\documentclass[useAMS,usenatbib]{mn2e}
\usepackage{graphicx}
\usepackage{times}

\begin{document}

\renewcommand{\topfraction}{1.0}
\renewcommand{\bottomfraction}{1.0}
\renewcommand{\textfraction}{0.0}

\title[Statistics of triple and quadruple stars]
{Comparative statistics and origin of triple and quadruple stars}

\author[Tokovinin]{A.~Tokovinin\thanks{E-mail:
atokovinin@ctio.noao.edu}\\
Cerro Tololo Inter-American Observatory, Casilla 603, La Serena, Chile}

\date{-}

\pagerange{\pageref{firstpage}--\pageref{lastpage}} \pubyear{2007}

\maketitle

\label{firstpage}

\begin{abstract}
The statistics of catalogued quadruple stars consisting
of two  binaries (hierarchy 2+2)  is studied in comparison  with triple
stars,   with  respective  sample   sizes  of   81  and   724.   Seven
representative quadruple systems are  discussed in greater detail. The
main conclusions are:
(i) Quadruple  systems of $\varepsilon$~Lyr type with similar
masses and inner  periods are common, in 42\% of  the sample the outer
mass ratio is  above 0.5 and the inner periods differ  by less than 10
times.
(ii) The distributions of the inner periods in triple and quadruple stars
are similar and bimodal. The inner mass ratios do not correlate with
the inner periods. 
(iii)  The statistics of  outer periods  and  mass ratios  in
triples and quadruples  are different. The median outer  mass ratio in
triples is 0.39  independently of the outer period,  which has a smooth
distribution.  In  contrast,  the  outer  periods  of  25\%  quadruples
concentrate in the narrow range from 10\,yr to 100\,yr, the outer mass
ratios of these tight quadruples are above 0.6 and their two inner periods
are similar to each other.
(iv) The outer and inner mass ratios in triple and quadruple stars are
not mutually correlated. In 13\% of quadruples both inner mass ratios
are above 0.85 (double twins). 
(v) The inner and outer  orbital angular momenta and periods in triple
and  quadruple  systems  with  inner  periods above  30\,d  show  some
correlation, the  ratio of outer-to-inner periods  is mostly comprised
between  5 and $10^4$.  In the  systems with  small period  ratios the
directions of the  orbital spins are correlated, while  in the systems
with large ratios they are not.
The properties of multiple stars  do not correspond to the products of
dynamical decay  of small clusters,  hence the N-body dynamics  is not
the  dominant  process  of   their  formation.   On  the  other  hand,
rotationally-driven  (cascade)   fragmentation  possibly  followed  by
migration  of  inner and/or  outer  orbits  to  shorter periods  is  a
promising  scenario to  explain  the origin  of  triple and  quadruple
stars.
\end{abstract}

\begin{keywords}
stars:formation -- stars:statistics -- binaries:general -- binaries:close
\end{keywords}

\section{Introduction}
\label{sec:intro}

Formation of binary and multiple stars is a subject of active research
and  debate,  still  remaining   one  of  the  major  unsolved  issues
\citep{IAU200}.  Reliable data on the statistics of binary and multiple
stars  in different  environments  are essential  for comparison  with
theory and further progress.  So  far, binary stars have received some
attention,  but  the  statistics  of higher  multiplicities  is  still
uncertain.   The aim  of this  paper is  to review  the  statistics of
multiple stars, focusing on quadruple systems, and to relate it to the
formation scenarios, whenever possible.

Multiple stellar systems are  often considered individually, hence the
formation  of each  object can  be explained  by a  unique  chain of
events. Yet, multiples  are not rare ``freaks'' in  the stellar world.
At  least  8\% of  solar-type  stars  have  three or  more  components
\citep{Tok08}.  The  nearest star, $\alpha$~Cen,  is triple.  Multiple
systems are normal products  of star formation, and the star-formation
theory will eventually be able to model their statistical properties.

A textbook example  of a quadruple system is  $\varepsilon$~Lyr -- two
wide  visual binaries  on a  still  wider common  orbit. Two  striking
properties of this system are the similarity of the masses of all four
components and comparable periods of the inner sub-systems, suggesting
that the  formation process was  not completely random.  We  show that
such quadruples  with similar components and inner  periods are indeed
typical.

One reason why  the statistics of multiple stars  remains murky is the
difficulty of  accounting for the  observational selection. Catalogued
multiples  result from  random  discoveries using  a  wide variety  of
techniques, so  quantifying the observational bias  seems hopeless.  A
better  approach would  be to  study volume-limited  samples,  but the
number of  multiples even  among nearby G-dwarfs  is small,  while the
incompleteness of  higher-order hierarchies is  still significant. So,
at present,  the only way to  look at the statistics  of $N>2$ stellar
systems  is  by   using  catalogues.   \citet{Fekel81}  pioneered  the
statistical approach  to triple stars.  The number  of known multiples
has significantly increased since then, making it possible to take the
first look at  systems with more than three stars.   About half of the
known quadruples were discovered during  the last 20 years as a result
of  systematic radial velocity  programs and  high-resolution imaging,
the  latest ones  \citep{Shkolnik}  are not  included  in the  present
study.

Here  we use  the updated  Multiple Star  Catalogue \citep[][hereafter
MSC]{MSC} to study  the statistics of quadruple stars  composed of two
close  pairs orbiting each  other.  The  problem of  the observational
selection is partially addressed by comparing with the triple systems,
assuming  that  the  selection  effects  in both  cases  are  similar.
Although the exact form of this selection filter remains undetermined,
we can make  reasonable assumptions about its behaviour  to figure out
which statistical features are real.   A catalogue such as MSC gives a
distorted,  but  still  useful  information  on  the  true  underlying
statistics of multiple stars.

We begin with a short  review of multiple-star formation processes and
their  predictions for  the  statistics (Sect.~\ref{sec:form}).   Some
prototypical  multiple stars are  presented in  Sect.~\ref{sec:sys} to
introduce relevant terminology and to give a feeling of the data.  The
samples   of   quadruple   and   triple   stars   are   described   in
Sect.~\ref{sec:samp}  together   with  the  selection   effects.   The
comparative   statistics   is   presented   in   Sect.~\ref{sec:comp}.
Section~\ref{sec:sum}  summarises  our   findings  and  discusses  the
implications for multiple-star formation.

\section{Formation processes}
\label{sec:form}

A short review of multiple-star formation scenarios is given to put
the statistics  in proper context.   The sequence of events,  from the
fragmentation of pre-stellar cores to the N-body interaction and orbit
migration, seems well established. Still, theoretical predictions vary
widely,  depending  on  the  initial  conditions,  physical  processes
involved or modelled, and computational details. This is a rapidly
evolving field. 

{\bf Fragmentation}  is believed to be the  dominant mechanism forming
binary                and                multiple                stars
\citep{Bonnell2000,Bate02,DD04,Goodwin04,Machida08}.
\citet{Bodenheimer78} envisioned a {\em cascade fragmentation} where a
rotating  core   first  collapses  into  a  ring   or  disk  supported
centrifugally.  The  ring fragments, and,  if the angular  momentum of
the fragments is high,  they undergo further rotational fragmentation,
forming a 2+2 hierarchical quadruple.

Real  pre-stellar cores  are  in the  state  of complex  ``turbulent''
motions, so the cascade fragmentation gives an over-simplified picture
which, nevertheless, captures the essential physics.  A collapse of an
isolated  turbulent  core of  gas  initially  creates filaments,  then
clumps in  the filament (or inter-sections of  several filaments) form
stars  or  sub-systems  \citep[e.g.][]{Krumholz07,Goodwin04}.   As  an
alternative  to the  rotationally-driven  fragmentation, {\em  prompt}
fragmentation  is induced by  an external  shock wave  or by  an $m=2$
gravitational  perturbation  from  another  companion.   The  Larson's
velocity-size relation  $V \propto R^{0.5}$ means  that in pre-stellar
cores  most of  the  specific  angular momentum  $j$  is contained  in
large-scale motions,  $j \propto RV  \propto R^{1.5}$. If the  core is
isolated, this momentum  will be conserved in the  orbital motion of a
wide binary or an outer sub-system.

Simulations  show  that  accretion   disks  can  also  produce  binary
companions  by  fragmentation  and  can  even  form  multiple  systems
\citep[cf.  Fig.~4  in][]{Bonnell2000}.  \citet{Stamatellos07} studied
disk fragmentation into several low-mass components, some of which are
subsequently ejected.   Unlike the first generation  of stars produced
directly  by  fragmentation, disk-generated  companions  form over  an
extended  period of  time, during  the whole  episode of  collapse and
accretion.    \citet{DD04}  show,  however,   that  disk-fragmentation
companions have  less chance  to accrete as  much matter as  the first
stars and are often ejected during subsequent dynamical evolution.

{\bf  Accretion.}    Binary  systems  produced   by  fragmentation  at
isothermal  stage of  collapse  are relatively  wide,  from $10^2$  to
$10^4$  AU  \citep[cf.   the  scaling formula  in][]{Sterzik03}.   The
fragments continue to accrete surrounding gas, so the final parameters
of  a  binary (or  a  sub-system)  are  determined by  the  accretion.
\citet{Bate2000}  studied the  evolution  of the  separation and  mass
ratio of an accreting proto-binary  under the assumption that there is
no angular momentum transport in the accreted gas.  The results depend
on the density and rotation profiles of the cloud.  Typically (but not
always) the mass  ratio of an accreting binary  increases and tends to
unity, while the orbital period decreases.  This trend is confirmed by
numerous hydrodynamical  simulations.  Binaries with  nearly identical
components,  {\em  twins}  \citep{Halbwachs03,Lucy06,Sod07}, are  most
naturally  explained  by  accretion.   Fragmentation  occurs  also  at
adiabatic  stage of  the collapse  and at  higher  densities, creating
close binaries \citep{Machida08}.  However, in this case the fragments
have very low masses.  If these fragments do not merge, the parameters
of the resulting close binaries are still settled by the accretion.

{\bf N-body  dynamics} plays a  role in a non-hierarchical  cluster of
proto-stars that is  formed when several fragments fall  to the common
mass centre of a pre-stellar core. The cluster disintegrates, ejecting
some stars  and leaving  a stable binary  or multiple  system, usually
with the most massive stars  paired together. As the free-fall time of
stars  and gas  to  the centre  is  the same,  accretion and  dynamics
proceed  simultaneously.   Pure  N-body  dynamics  of  small  clusters
\citep{SD98} produces  unrealistic multiples with  small period ratios
and  large outer eccentricities  \citep[cf.  Fig.~6  in][]{ST02}.  The
presence of gas  does change the character of  the process compared to
the point-mass  dynamics.  Extensive simulations  of fragmenting cores
including  accretion   and  dynamical  evolution   were  performed  by
\citet{DD04}.   They find  that the  chaotic dynamics  disrupts mostly
wide  and/or low-mass companions  with low  binding energies,  but has
little effect on the  inner sub-systems with massive components formed
by fragmentation  and accretion.  Component masses  in these surviving
multiples are comparable,  as in the real systems  (see below). In the
10 simulated  runs, three 2+2  quadruples were formed  within 0.5\,Myr
from  the  start  of  the collapse,  two of  them  with  additional  outer
components.   All  those   quadruples  survived  subsequent  dynamical
evolution of the mini-clusters.

Many  authors \citep[e.g.][]{Sterzik03,Goodwin04} believe  that N-body
interaction is  a necessary  step in the  formation of  multiple stars
because  it  helps  to  reduce  orbital separations  by  an  order  of
magnitude  compared  to the  fragmentation  scale.   We question  this
assertion.   As large-scale  motions of  the core  contain significant
angular  momentum, the  fragments  do not  necessarily  fall onto  the
centre and  interact dynamically,  but rather may  end up in  a stable
wide binary, possibly containing  inner sub-systems.  This scenario is
conceptually close to  the cascade fragmentation and it  can produce a
hierarchical 2+2 quadruple  system like $\varepsilon$~Lyr with similar
components' masses and inner  orbits \citep{DD04}.  The orbital scales
of  the  $\varepsilon$~Lyr system roughly  match the  initial  fragmentation
scales, so  no further orbit evolution is  required.  The eccentricity
of the outer  orbit in this scenario should  be moderate, the multiple
system is hierarchical and dynamically stable.

{\bf  Orbit migration}  is an  accepted theory  for  ``hot Jupiters'',
where the  orbit of a planet  shrinks by interacting  with the gaseous
disk.  It is necessary to  evoke some kind of migration for explaining
the  origin of  close binaries,  unless  they are  formed directly  by
fragmentation  at high  densities.  The  eccentricity  distributions of
exo-planets   and  spectroscopic   binaries  are   strikingly  similar
\citep{Ribas07},  suggesting  that   interaction  with  disks  may  be
important in  both cases.   Unlike planets, stellar  components cannot
migrate  in low-mass  debris  disks, but  massive  accretion disks  at
earlier   stages  could  act   in  a   similar  way   \citep[type  III
migration,][]{Mig3}.   Accretion onto  a  binary also  can reduce  its
period,    even    without     any    angular    momentum    transport
\citep[e.g.][]{Bate2000}, but  it is likely that in  this case massive
disks are formed as well and affect the orbit of the binary.

There  may be  several  mechanisms of  orbit  migration. Their  common
characteristic is that the  orbital angular momentum is transferred to
some  other  body,  while  the  potential  energy  released  from  the
component's  approach is  dissipated. Apart  from the  disk migration,
candidate processes are interactions  with a jet \citep{Reipurth04} or
with   a   magnetic   wind.    Kozai  cycles   with   tidal   friction
\citep[KCTF,][]{Eggleton}  can  be viewed  as  yet  another flavor  of
migration  where the  angular momentum  is deposited  into  a tertiary
component while the orbital energy is dissipated by tides.

Statistics  of   close  binaries   help  to  quantify   the  migration
observationally.  It  has been established \citep{Tok06}  that not all
low-mass spectroscopic  binaries (SBs)  with periods below  30\,d have
tertiary  components,  so  these  SBs  are  produced  by  a  migration
mechanism  distinct from the  KCTF. The  SBs with  tertiary companions
have  shorter   periods  than  pure  binaries,  but   the  mass  ratio
distributions of  these two  sub-populations are identical,  hence the
dominant migration process  does not change the mass  ratio.  Many SBs
have  tertiary components  too distant  to cause  KCTF  migration, yet
their  periods are, statistically,  shorter than  the periods  of pure
SBs, so  the migration is  likely enhanced by  the presence of  even a
distant companion.

The  properties  of  multiple  stars  are reviewed  with  the  aim  to
establish which formation processes are dominant.

\section{Typical quadruple systems}
\label{sec:sys}

\begin{figure}
\includegraphics[width=8.5cm]{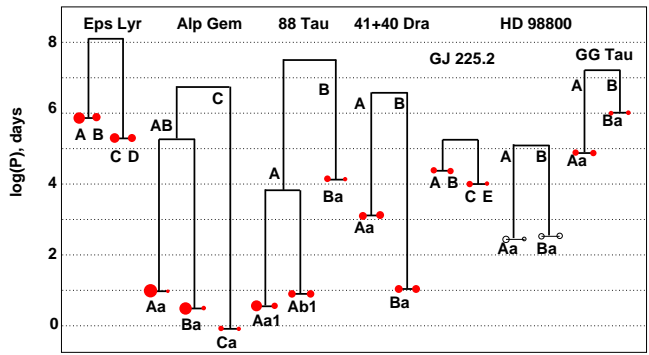} 
\caption{Hierarchy of  the selected  quadruple stars.  The  periods of
sub-systems are shown by the horizontal lines, the size of the circles
depicting  stellar   components  is  roughly   proportional  to  their
masses. Note that  the ratios of outer to  inner periods are typically
large.  
\label{fig:h}}
\end{figure}

The data  for this  study are  taken from the  current version  of the
Multiple  Star Catalogue  (MSC) \citep{MSC}.   We begin  by presenting
seven typical quadruple systems to  illustrate the nature of the input
data and its reliability.  Figure~\ref{fig:h} depicts the hierarchy of
these systems, where a  hierarchical multiple system is represented as
a combination  of elementary binaries composed either  of single stars
or of closer  pairs.  Relevant parameters of the  systems are gathered
in     Table~\ref{tab:1},      with     notations     explained     in
Sect.~\ref{sec:notation}.   All systems  are identified  by  WDS codes
based on the 2000 coordinates \citep{WDS}, with additional identifiers
provided  in the  2-nd column.   We  focus only  on quadruple  systems
composed of two close pairs on a common wide orbit (hierarchy 2+2) and
leave  aside quadruples  of  hierarchy (2+1)+1,  i.e.  a  hierarchical
triple orbited  by a distant  companion.  Some triples  and quadruples
considered  here  contain  additional  components, as  illustrated  by
Fig.~\ref{fig:h}.   The   terms  {\em  triple}   and  {\em  quadruple}
generally mean systems containing {\em  exactly} 3 or 4 stars. In this
paper we use them to denote  objects from our samples, so some systems
called here ``triple'' contain more than 3 stars.

\begin{table*}
\center
\caption{Selected quadruple systems}
\label{tab:1}
\begin{tabular}{l l ccc | cccc }
\hline
WDS(2000) & Name & \multicolumn{3}{c|}{Periods }  & \multicolumn{4}{c}{Masses,
  $M_\odot$}  \\
 & &  $P_L$ & $P_{S1}$ &  $P_{S2}$ &  $M_{1,1}$  & $M_{1,2}$ &  $M_{2,1}$ &  $M_{2,1}$   \\
\hline
18443+3938 & $\varepsilon$ Lyr & 340 ky &1804 y &585 y& 2.31 a & 1.62 a & 1.86 a & 1.70 a  \\
07346+3153 & $\alpha$ Gem     & 445 y  & 9.2 d & 3.0 d  &  2.76 a & 0.47 m & 2.98 a & 0.24 m \\ 
             &                & 14 ky & 445 y & 0.8 d    & 3.22 s & 3.23 s & 0.59 * & 0.58 *  \\      
04356+1010 & 88 Tau           & 18 y & 3.6 d & 7.9 d           & 2.06 * & 1.36 * & 1.07 * & 1.06 * \\ 
       &                      & 80 ky & 18 y & 3.7 y           & 3.42 s & 2.13 s & 1.18 a & 0.15 m  \\
18002+8000 & 41+40 Dra        & 10.3 ky & 1247 d & 10.5 d & 1.39 * & 1.30 * & 1.32 * & 1.20 *  \\
06003$-$3103 & GJ 225.2       & 460 y & 67.7 y & 23.7 y  & 0.67 * & 0.52 * & 0.69 * & 0.20 *  \\
11221$-$2447 & HD 98800       & 345 y & 262 d & 315 d        & 0.80 * & 0.67 q & 0.90 * & 0.19 m   \\
04325+1731 & GG Tau           & 45 ky & 200 y & 2700 y         & 0.78 * & 0.68 * & 0.11 * & 0.04 * \\
\hline
\end{tabular}
\end{table*}

{\bf Epsilon Lyrae} is a visual quadruple composed of four A-type Main
Sequence (MS) stars arranged in two  pairs AB and CD.  The component C
has been  resolved twice with  speckle interferometry into  a $0.19''$
binary  Cab=CHARA 77,  but the  existence of  this  sub-system remains
doubtful because  of many negative  observations.  So, we  ignore this
uncertain couple  Cab. However,  some  components  believed  now to  be
single stars may be resolved in the future.


{\bf  Alpha  Geminorum}  is  a  classical  multiple  star  where  both
components A and  B of a 445-yr visual  binary are close spectroscopic
pairs with A-type  primaries and low mass ratios.   The eclipsing pair
of  red  dwarfs YY~Gem  (component  C)  at  angular distance  $72.5''$
(projected distance  1145\,AU) also belongs  to the system,  making it
sextuple. Thus,  there are 2  quadruples of (2+2) hierarchy  here: one
with the outer  445-yr orbit AB and another with  the outer orbit AB-C
and the  inner sub-systems  Cab and AB.   In the following,  we ignore
{\em composite}  quadruples, like the  outer one in  $\alpha$~Gem, and
consider only {\em simple} quadruples  composed of just 4 stars (with,
possibly, outer components), like the  inner system AB. In the case of
$\alpha$~Gem, we  can be sure  that there are  only 4 stars in  the AB
system, but our  sample of simple quadruples can  contain some systems
with yet undiscovered sub-structure.

{\bf  88  Tau}  is  a  sextuple  system with  the  same  structure  as
$\alpha$~Gem. It is cited here to illustrate that such systems are not
exceptional.  The  projected distance  between the outer  components A
and B  is 3500\,AU,  with A being  quadruple and B  binary.  Recently,
interferometric  astrometry has permitted  to effectively  resolve the
inner  sub-systems   Aa1,Aa2  and  Ab1,Ab2  and   to  determine  their
orientation  with respect  to  the 18-yr  orbit Aab  \citep{Lane2007}.
Interestingly,  the  inclination  of  the more  massive  inner  binary
Aa1,Aa2 relative  to the  outer orbit was  found to be  $143^\circ \pm
2.5^\circ$, i.e.   it is  {\em counter-rotating}.  The  inclination of
the second sub-system Ab1,Ab2  is either $82^\circ$ or $58^\circ$, its
orbit is less secure.

{\bf 41+40 Dra} is a  well-studied quadruple with F-type MS components
\citep{41Dra}.   A combination of  observational constraints  makes the
existence  of any  additional companions  extremely unlikely,  so this
system is a genuine quadruple.  The mass ratios of the sub-systems Aab
and Bab  are nearly the same, 0.93  and 0.91.  The periods  of Aab and
Bab are  different, but the orbit  of Aab is highly  eccentric, so the
{\em angular momenta} of Aab and Bab are similar. The pairs Aab and AB
rotate in opposite directions as projected onto the sky.  Although the
outer orbit  is still undetermined, we  know that it  is not co-planar
with  the  orbit Aab,  the  probable  relative  inclination is  either
$60^\circ$ or $160^\circ$.

{\bf GJ  225.2} was a  visual triple star  until recently, when  a new
sub-system CE has been  discovered with adaptive optics \citep{GJ2252}.
The period of CE was estimated from the astrometric  perturbations. So far,
there  was no  radial  velocity  (RV) monitoring  of  these stars,  so
additional  close components  may  still hide  inside.  The system  is
apparently  old, as  it has  a high  velocity relative  to the  Sun. A
premature  orbit of the  outer system  AB-C with  $P_L =  390$\,yr was
published, but all we can say now is that the outer period is of the
order of 500\,yr and that all three orbits could be close to
co-planarity.

{\bf  HD  98800}  is   a  young,  pre-Main  Sequence  (PMS)  quadruple
consisting of  two spectroscopic  binaries Aab and  Bab at  $2''$ from
each  other.   The  sub-system  Bab was  resolved  by  interferometry,
providing  its ``visual''  orbit  and a  reliable  measurement of  the
masses \citep{Boden05}.   The outer orbit  of AB is  only preliminary,
but  there  are indications  that  the AB  and  Bab  orbits have  only
moderate mutual  inclination.  The mass of  Ab is still  a lower limit
from the single-lined spectroscopic orbit.  The Bab sub-system is more
massive  than  Aab  and  is  surrounded  by  a  massive  disk  causing
extinction and flux variability, whereas Aab has no such disk.

{\bf GG  Tau} is another PMS  quadruple only 1--2\,Myr  old.  All four
components   are  resolved,   the  masses   are  estimated   from  the
evolutionary  tracks,   with  Bb  possibly   being  sub-stellar.   The
sub-system  Aab  is  surrounded  by  a  ring-like  disk,  there  is  a
circum-quadruple  disk   as  well.   \citet{Beust06}   used  the  disk
parameters to constrain the  dynamics of this quadruple. They conclude
that the  relative inclinations between  Aab, its disk, and  the outer
orbit   are   likely  moderate,   so   the   system  is   more-or-less
co-planar. Interestingly,  the pair Bab is  close to the  limit of the
dynamical stability, and, if  stable, it is probably counter-rotating.
GG~Tau may  disintegrate during further  dynamical evolution, ejecting
Bb and leaving only a triple system.

\begin{table}
\center
\caption{Sample of triple systems (fragment)}
\label{tab:2}
\begin{tabular}{l c c c c c c}
\hline
WDS(2000) & $\log P_L$ &  $\log P_S$ & $M_1$ & $M_2$ & $M_3$ & Plx \\ 
          & day        & day         & $M_\odot$ & $M_\odot$ &
$M_\odot$ & mas \\
\hline
00046-4044 &  6.33 & 3.10 & 0.06 a & 0.06 * & 0.26 *  &  76.9 \\ 
00057+4548 &  7.85 & 5.74 & 0.65 a & 0.47 a & 0.37 a  &  85.1 \\ 
00063+5826 &  4.59 & 1.68 & 0.86 * & 0.31 * & 0.98 *  &  49.3 \\  
00125+1434 &  3.19 & 0.27 & 0.78 m & 0.54 a & 0.52 q  &  24.7 \\  
00150+0849 &  6.73 &$-$0.08 & 1.24 a & 1.15 * & 1.13 *  &  12.5 \\ 
\hline
\end{tabular}
\end{table}

\begin{table*}
\center
\caption{Sample of quadruple systems (fragment)}
\label{tab:3}
\begin{tabular}{l ccc  cccc c}
\hline
WDS(2000) & $\log P_L$ & $\log P_{S1}$ & $\log P_{S2}$  & $M_{1,1}$ &
$M_{1,2}$ & $M_{2,1}$ & $M_{2,2}$ &  Plx \\ 
          & day & day & day &  $M_\odot$ & $M_\odot$ & $M_\odot$ & $M_\odot$ & mas \\
\hline
00134+2659 &  7.20 & 5.13 & 0.78 & 2.50 a & 1.54 a & 0.92 * & 0.88 * &   8.1 \\
00247-2709 &  4.16 & 2.30 & 3.41 & 0.10 * & 0.08 * & 0.08 * & 0.08 * & 132.8 \\
00316-6258 &  6.68 & 5.26 & 4.21 & 3.84 a & 0.40 v & 2.76 a & 1.94 a &  23.4 \\
00345-0433 &  7.18 & 5.56 & 1.91 & 1.70 * & 1.40 * & 1.14 * & 0.10 m &  10.0 \\
00364-4908 &  8.57 & 0.57 & 1.34 & 0.98 a & 0.83 a & 0.78 a & 0.31 m &  22.8 \\
\hline
Notes: \\	   	 
\multicolumn{9}{l} {00134+2659 = HR 40. VB (368y G0III) and SB (6d G7V) at  18$''$. } \\
\multicolumn{9}{l}{00247-2709 =  GJ 2005, visual  quadruple composed
  of  M5 dwarfs, 0.27$''$ and 0.05$''$  at 1.07$''$ from each  other. }\\
  &\multicolumn{8}{l}{Short estimated  periods, but no
orbits yet.} \\
\multicolumn{9}{l}{00316-6258 = HR 126+127+125, Beta Tuc, 6 components. The inner
quadruple contains 2 VBs (2.4$''$ B9V and 44.7y A2V)}\\ 
  &\multicolumn{8}{l}{at 27$''$. The distant pair at 544$''$ is itself a 0.1$''$ VB with a short, but yet unknown period.}  \\
\multicolumn{9}{l}{00345-0433 = HIP 2713. VB (1.9$''$ G8II) and SB1 (81d G0V) at 19.5$''$.}\\
\multicolumn{9}{l}{00364-4908 = HIP 2888 = GJ 24AB. Two SBs (3.7d G3V  and 22d K0V) at 329$''$, CPM. } \\
\end{tabular} 
\end{table*}

\section{Statistical samples}
\label{sec:samp}

\subsection{Parameters and notation}
\label{sec:notation}

Hierarchical  multiple systems  are  described by  a  large number  of
parameters  --  component's  masses   and  orbital  elements  of  each
sub-system. We concentrate  on masses and periods as  most relevant to
the formation processes and touch on other characteristics (eccentricity,
mutual orbit inclination) only briefly. The semi-major axis is related
to  the  period  by  the  third  Kepler's  law  and  is  statistically
equivalent to period, except for a weak dependence on the total mass.

Consistent notation of the  parameters is required to avoid confusion.
For {\em triple stars}, the outer (long) and inner (short) periods are
denoted as $P_L$ and $P_S$,  the masses of the inner binary companions
as $M_1$ and $M_2$ and the mass of the distant tertiary as $M_3$.  The
inner and outer mass ratios $q_S$ and  $q_{L3}$ are defined as
\begin{equation}
 q_S = \frac{M_2}{M_1}, \;\;\;\;  q_{L3} = \frac{M_3}{M_1 + M_2},
\label{eq:q3}
\end{equation}
although an alternative definition $q_{L3}' = M_3/M_1$ is also possible. 
The outer mass ratio 
can take values $q_{L3}>1$ if the tertiary is more massive than the
inner binary (this happens  in 5\% of the triples). 

The  notation for quadruple  stars is  as follows:  the period  of the
outer system  is $P_L$, $P_{S1}$ and  $P_{S2}$ are the  periods of the
inner sub-systems, with  index 1 referring to the  sub-system with the
largest total mass.  The masses  of all 4 components are designated as
$M_{1,1}$, $M_{1,2}$, $M_{2,1}$, and $M_{2,1}$.

The mass ratios of quadruple stars can be defined in various ways. We
use the following definitions: 
\begin{equation}
q_{S1} = \frac{M_{1,2}}{M_{1,1}}, \;\;\;\;\; q_{S2} = \frac{M_{2,2}}{M_{2,1}},
\;\;\;\;\;  q_{L4} = \frac{M_{2,1}+ M_{2,2}}{M_{1,1} + M_{1,2}} .  
\label{eq:q}
\end{equation}
One  can argue that  the outer  mass ratios  for triple  and quadruple
stars are not directly comparable because the latter
compares two  stellar pairs.  An  alternative definition of  the outer
mass ratio in quadruple stars can be $q_{L4}'$,
\begin{equation}
q_{L4}' = \frac{M_{2,1}}{M_{1,1} + M_{1,2}} = \frac{q_{L4}}{1 + q_{S2}}.
\label{eq:q4}
\end{equation}
This  parameter can  be  compared with  $q_{L3}$  more directly.   For
example, when all components have  equal masses, we have $q_{L4} = 1$,
but $q_{L4}'  = q_{L3} =  0.5$. The choice  of the most  relevant mass
ratio  definition  depends  on  the implied  formation  process,  e.g.
$q_{L3}$ and $q_{L4}$ can  characterise the fragmentation of the outer
sub-system.

The  data on  the periods  and  masses are  taken from  the MSC.   The
periods are either  known more or less exactly  from the spectroscopic
or visual  orbits or, otherwise, estimated roughly  from the projected
separations in the case of  wide binaries.  The components' masses are
estimated  by  various techniques,  as  reflected  by the  1-character
codes.  The most common mass codes are {\bf a} (mass from the spectral
type), {\bf *} (mass  estimates from orbits, isochrone fitting, etc.),
{\bf  q} (secondary  of  an SB2),  {\bf  m} (minimum  mass  of an  SB1
secondary),  {\bf v}  (estimate  from the  magnitude  difference in  a
visual   binary),  and   {\bf  :}   (uncertain  mass,   e.g.    in  an
interferometric binary  without relative photometry  where a magnitude
difference of  2 is assumed).   The mass code  {\bf s} means  that the
``component'' consists of several stars and the sum of their masses is
given.   The   reader   is   referred   to   the   MSC   for   further
details. Obviously, the mass estimates  in the MSC are quite crude, so
the  mass ratios discussed  below are  only indicative  of statistical
trends, rather than well-measured.

\subsection{Triple and quadruple systems}

The  data on triple  and quadruple  stars are  listed in  the Tables~2
and~3,   respectively.    These   tables   are   available   in   full
electronically,  their fragments are  given  here.   Each system  is
identified by  its WDS~(2000) code.  Additional  identifiers are given
in  the Notes  to Table~3.   More information  on each  system  can be
obtained                from                the                on-line
MSC\footnote{http://www.ctio.noao.edu/\~{}atokovin/stars/index.php}.

We selected  all ``simple'' triple  systems composed of an  inner pair
and a distant  single tertiary, with possibly some  other more distant
components.  The  systems with uncertain masses (code  :) are omitted,
leaving $N=724$ triples listed in Table~2.

\begin{figure}
\includegraphics[width=8cm]{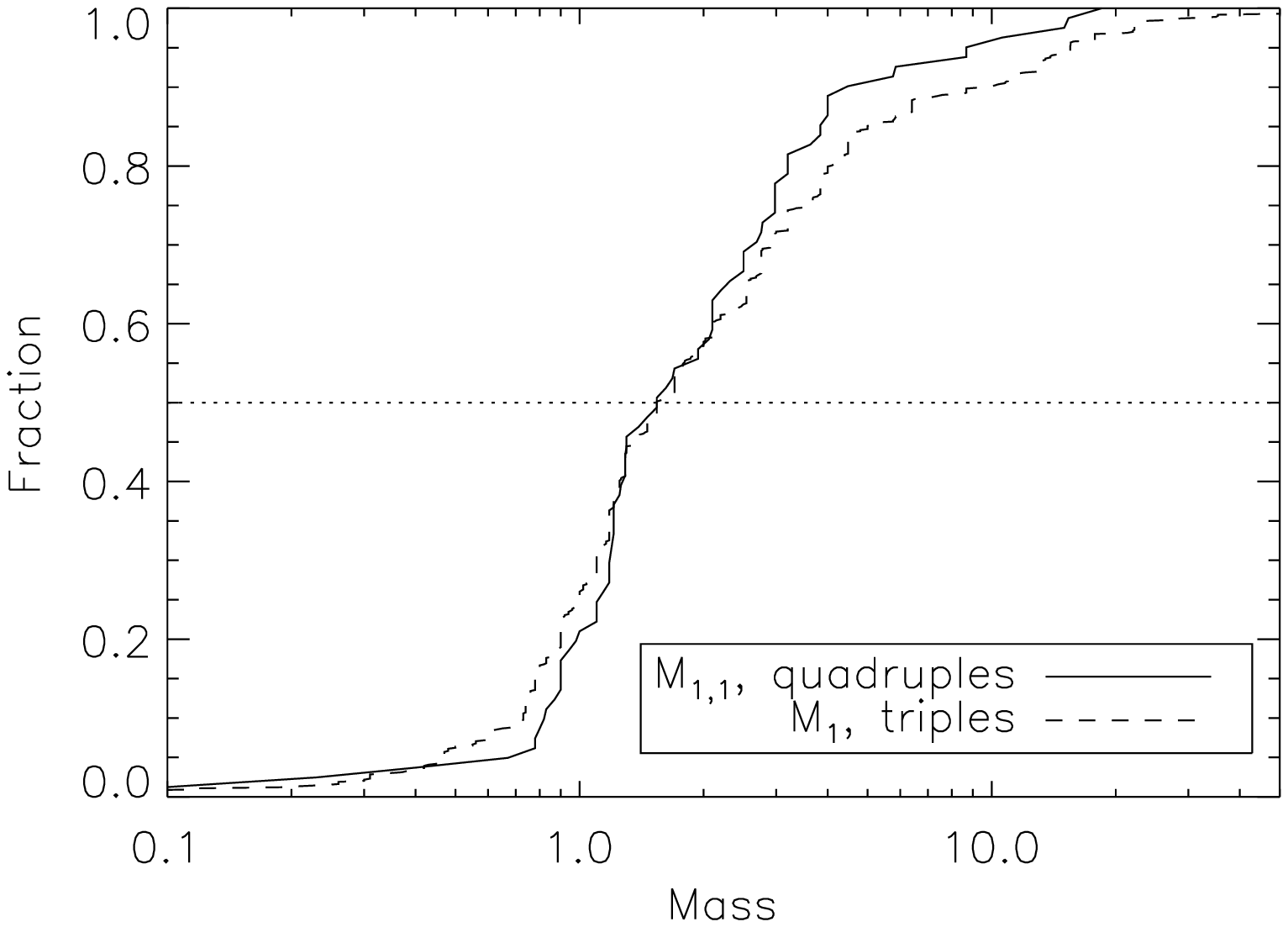}
\includegraphics[width=8cm]{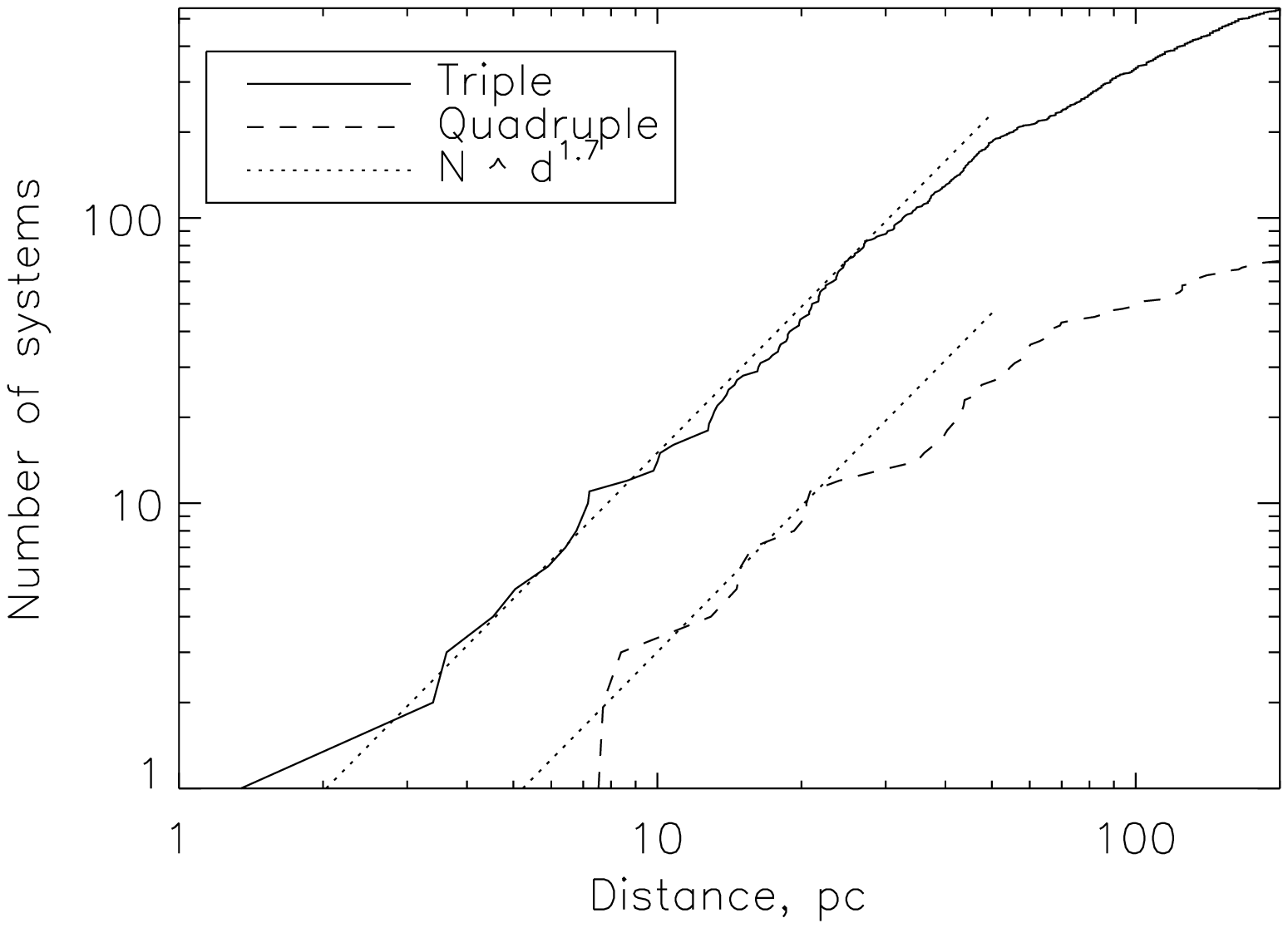}
\caption{Top: Cumulative distributions of  the primary masses: $M_{1,1}$ in
81 quadruples (full line) and $M_1$ in 724 triples (dashed line).
Bottom: The number of triple and quadruple systems within given distance
$d$. The dotted lines show the $d^{1.7}$ trends with coefficients 0.3
and 0.06. 
%
\label{fig:m1hist}}
\end{figure}

The sample  of 2+2 quadruples from  the MSC contains  135 entries.  We
exclude  from the statistical  analysis composite  quadruples, systems
where  some  periods  or  masses  are  not  known  (e.g.   astrometric
sub-systems without computed orbits), leaving 81
simple quadruples.  Periods,  masses, and mass codes are  given in the
electronic Table~3, together with the notes on individual systems.  We
keep in  the sample  3 systems (04226+2538,  11395$-$6524, 17146+1423)
where one of  the 4 masses is poorly determined  (code :), but exclude
some  PMS multiples  with  yet uncertain  masses  like those  recently
discovered by  \citet{Correia}.  Three quadruple  systems from Table~3
(SZ~Cam, $\mu$~Ori, QZ~Car) belong to open clusters.

Both samples are quite heterogeneous with respect to masses, ages, and
reliability of the data.  The most massive is the 40~$M_\odot$ primary
component of the quadruple  QZ~Car (10444$-$6000), while the red dwarf
primary in the quadruple  GJ~2005 (00247$-$2709) has the smallest mass
of   0.1~$M_\odot$.   Figure~\ref{fig:m1hist}  shows   the  cumulative
distributions of the primary-component  masses in the samples of triple
and quadruple stars. In both cases about 80\% of primaries have masses
between 0.8  and 5~$M_\odot$, with the median  mass of $1.7\,M_\odot$.
We tried  to split the quadruples  into low- and  high-mass parts around
this median and found the statistics to be similar.

In a reasonably complete nearby  sample the number of objects within a
given  distance $d$  would be  proportional  to $d^3$.  The number  of
triple and  quadruple stars in  our samples is rather  proportional to
$d^{1.7}$ because  low-mass objects are located  closer than high-mass
ones (Fig.~\ref{fig:m1hist}, bottom).  Such  weak dependence on $d$ is
expected for  magnitude-limited samples.  We deduce from  the ratio of
the coefficients in  the $d^{1.7}$ fits that triples  are 5 times more
frequent than  2+2 quadruples.  We recover the $N \propto d^3$ law 
out to  $d< 40$\,pc  if  only primaries more massive  than 1\,$M_\odot$
are retained.  The median distances  to triples and quadruples are
100\,pc and 70\,pc, respectively.

\subsection{Selection effects}
\label{sec:sel}

Multiple  stars  are  discovered  by  different  techniques  or  their
combinations, each with  its own biases.  Moreover, the  choice of the
surveyed objects  is random rather  than systematic. As a  result, the
current  knowledge of  multiplicity is  incomplete even  in  the solar
neighbourhood.  Here we study the  {\em statistics of the catalogue}, not
the  statistics of  real  multiple stars  in  the sky.   

It  is  standard  practise  to  adopt  some  models  of  observational
selection  for  deriving true  statistics  from  catalogues or  biased
samples.   This approach  is justified  for some  well-defined samples
\citep[e.g.][]{Tok06},   but  it   is   problematic  for   compilative
catalogues resulting from random discoveries, such as the MSC.  A {\em
selection function} $f$ is the ratio of objects in the sample to their
true  number in the  same stellar  population.  Here  we intentionally
leave the  selection function  undefined and do  not attempt  to study
true  distributions   of  periods   and  mass  ratios.    However,  we
investigate the correlations between  the parameters of the catalogued
systems  and try  to figure  out  the influence  of selection  effects
qualitatively.  Two reasonable assumptions  are made.
%
(i) The discovery of  a sub-system depends mostly on its period
$P$ and mass ratio $q$, so, to the first order, the selection function
has the form $f(P,q)$.
%
(ii) The selection  function $f(P,q)$ is smooth,  without features favouring
or disfavouring particular periods or mass ratios.
The selection  function acts  as a filter  superposed on  the observed
distribution.   If there  are correlations  between the  parameters of
real multiple  stars, they  are likely  to be seen  in the  catalogue as
well. Sharp  features in the  real distributions will be  recovered in
the catalogue, too.  

The integral  of $f(P,q)$ gives  the {\em completeness} of  the sample
$f_0$.  The MSC certainly  misses many low-mass companions. However, a
bold assumption that  we see only the ``tip of  the iceberg'' and that
stars are surrounded by swarms of yet-undiscovered low-mass companions
meets  with  problems,  at  least  for nearby  dwarfs.   For  example,
\citet{Egg07} find  that only 15\%  of stars previously  considered as
single and  surveyed with adaptive optics have  low-mass companions at
separations  above  7~AU.   Precise   radial  velocities  of  93\%  of
``single''  dwarfs  are  stable  \citep{Nidever02}.  It  is  therefore
reasonable  to assume  that the  completeness  of the  MSC for  nearby
dwarfs is of the order $f_0 \sim 0.5$ or higher.

The same  techniques are used for discovering  {\em inner} sub-systems
in   quadruple  and   triple   stars,  so   the  selection   functions
$f(P_{S},q_{S})$  for triples  and  $f(P_{S1},q_{S1})$ for  quadruples
should be similar.  Unfortunately, we cannot assume that the selection
functions  $f(P_{L},q_{L3})$  for  triples and  $f(P_{L},q_{L4})$  for
quadruples are the  same.  Discovering a faint tertiary  can be easier
than finding  that this  tertiary is in  fact double (cf.   the system
GJ~225.2).  A systematic  study of  tertiary components  has converted
many of  them into additional  pairs, promoting triples  to quadruples
\citep{TS02}.   It has  been shown  that 30\%  of  tertiary companions
believed to be single are in fact binary.

As mentioned above,  only in few multiple systems we  can be sure that
all components are  actually known.  When a new  component is found in
the future  at some  intermediate level (e.g.   orbiting a  close pair
with a more distant and already known tertiary), our current knowledge
of the system's  structure is obviously wrong, as  the values of periods
and masses in the Tables 2 and 3 correspond to some other hierarchical
levels. When  a new component  is found in  a close orbit  around some
star, a  simple triple or quadruple  is converted into  a complex one.
Hopefully, such incompleteness affects  only a fraction of the systems
and the distortion of the statistics caused by these missing levels is
tolerably  small.  It  can be  reduced by  restricting the  samples to
nearby, better-studied systems. 

\section{Comparison between triples and quadruples}
\label{sec:comp}

Periods and masses  of the companions are determined  by the formation
and  early  evolution, therefore  we  study  the  statistics of  these
parameters. Nevertheless,  a triple is  characterised by 5  numbers (2
periods  and 3  masses),  a quadruple  --  by 7  numbers. Looking  for
correlations in  this multi-parameter  space is not  easy, so  we give
below only some plots which appear to be most relevant. We begin by
looking at the periods in the inner and outer sub-systems. Then the
inner and outer mass ratios are compared. Finally, we explore
correlations of the inner and outer mass ratios with respective
periods, the correlations between the orbital angular momenta, and the
properties of quadruple systems.  

\subsection{Outer and inner periods}

\begin{figure}
\includegraphics[width=8cm]{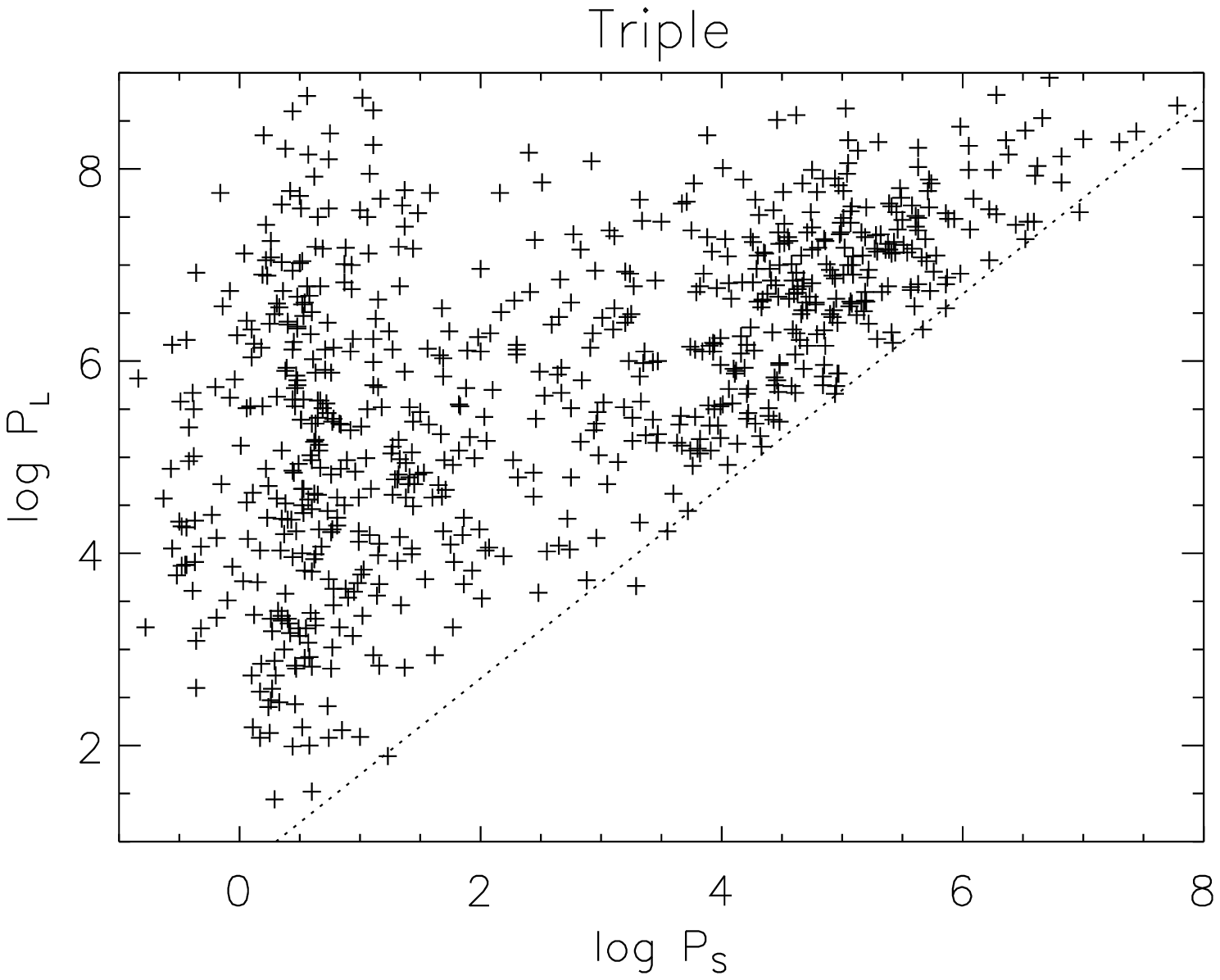}
\includegraphics[width=8cm]{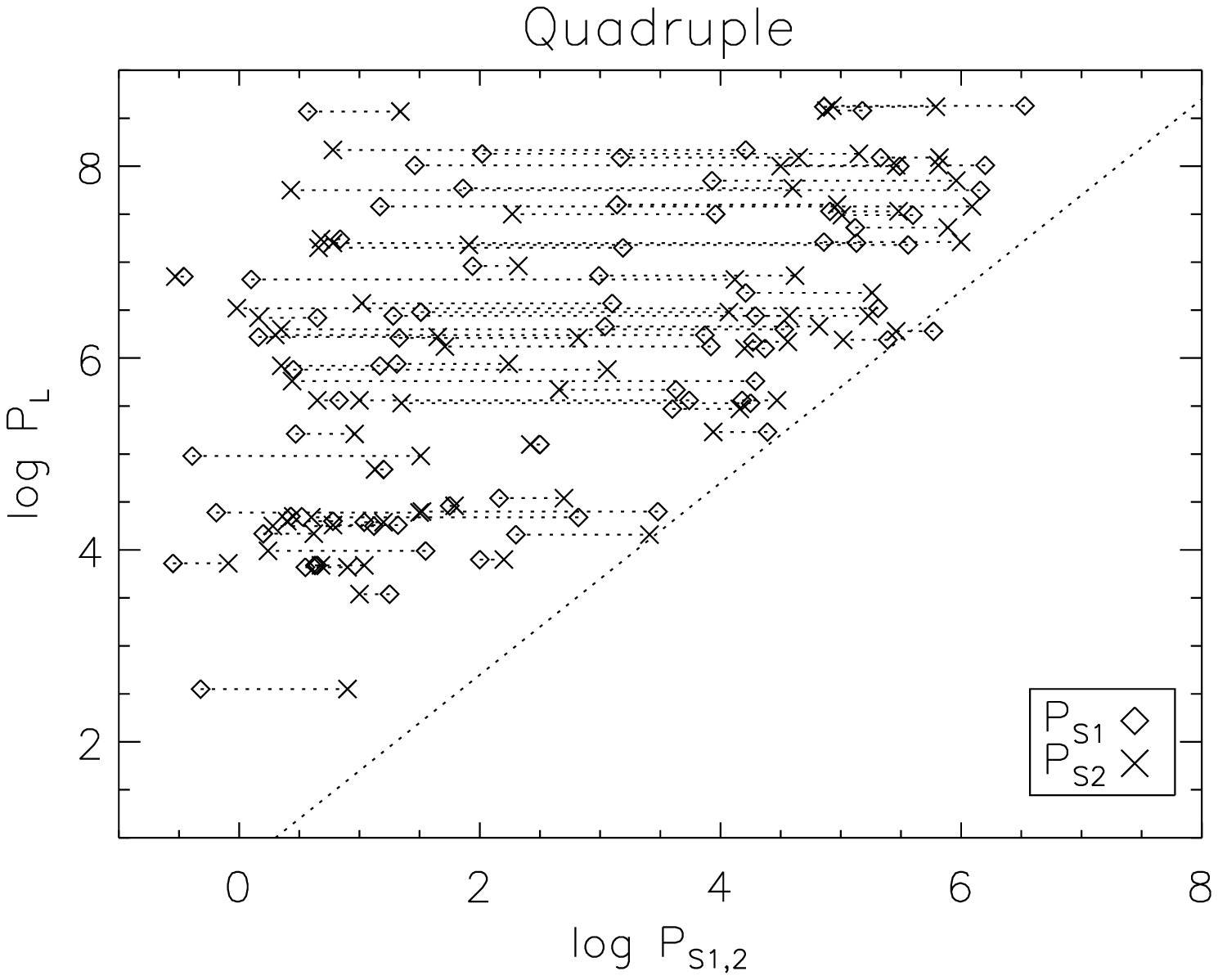} 
\caption{Comparison of the outer and inner periods in triple (top) and
quadruple (bottom)  stars.  All periods are in  days.  For quadruples,
$P_{S1}$  are plotted  as diamonds,  $P_{S2}$ as  crosses,  the points
belonging  to the  same system  are  connected by  dotted lines.   The
diagonal  lines delineate  the approximate  dynamical  stability limit
$P_L/P_S > 5$.
\label{fig:plps}}
\end{figure}

\begin{figure}                
\includegraphics[width=8cm]{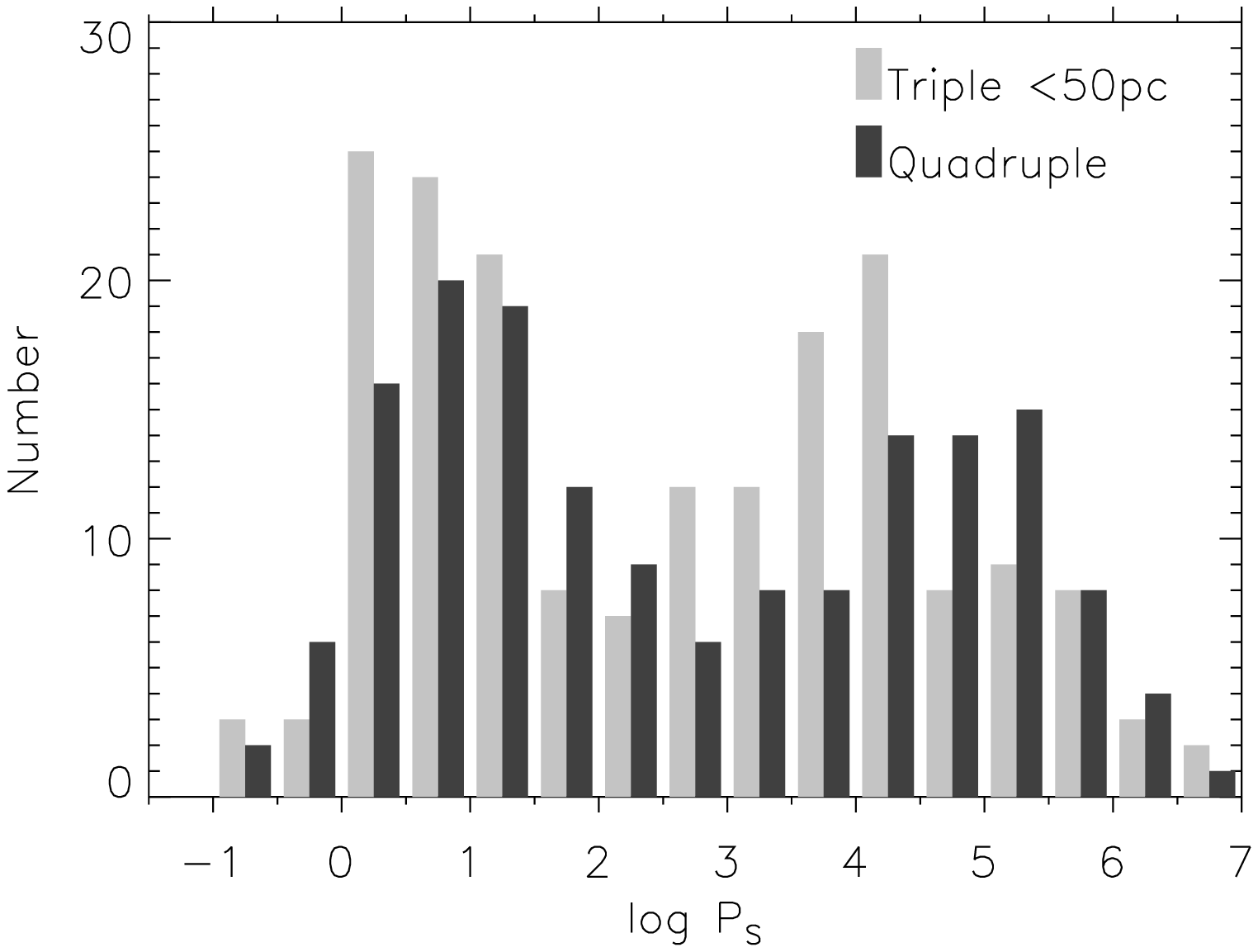}
\includegraphics[width=8cm]{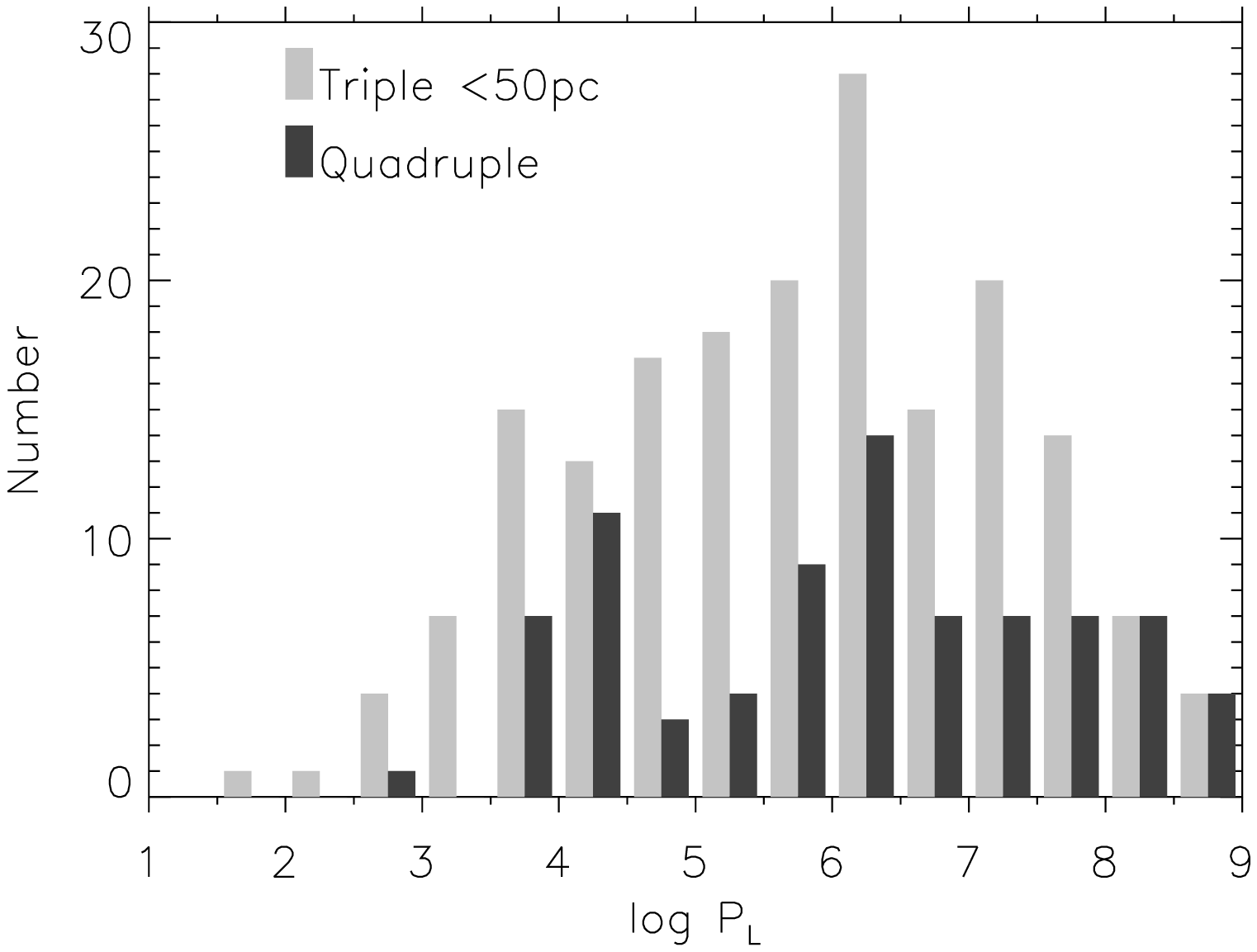} 
\caption{The  distributions  of the  inner  (top)  and outer  (bottom)
periods in triple  and quadruple stars.  The bars  refer to the period
bins of 0.5\,dex width and are displaced from the bin centers to avoid
overlap.  In quadruple stars, both inner periods are merged.
\label{fig:plpshist}}
\end{figure}

Figure~\ref{fig:plps}  plots   the  triples  and   quadruples  in  the
period-period diagram. Throughout the  paper, the periods are measured
in days and plotted on the logarithmic scale. Hierarchical systems are
dynamically stable  when $P_L/P_S >  5$ \citep[e.g.][]{Eggleton},
this  limit is  shown by  the  dashed lines.  As the  periods of  wide
systems  are estimated  only crudely,  some points  can  fall slightly
below this line.  The distributions  of inner and outer periods in 185
triple stars within  50\,pc and in all quadruple  stars are plotted in
Fig.~\ref{fig:plpshist}.

The {\em inner} periods in both triples and quadruples are distributed
in the same way, clustered in  two groups: close ($\log P_S <1.5$) and
wide ($\log P_S > 4$), with a partially filled gap in-between.  As the
discovery  techniques  for  the  close sub-systems  are  predominantly
spectroscopic and for the wide sub-system mostly visual, the gap could
be attributed to the selection effect, but it could also be real. Note
that the inner and outer periods in triple stars show some correlation
for $P_S > 30$\,d.  Most points are located in the band $5 < P_L/P_S <
10^4$ \citep[see  also][]{Tok04}, while the  space above this  band is
almost  empty, contributing  to the  gap in  the distributions  of the
inner periods.  This correlation must be genuine because the inner and
outer  sub-systems are  discovered  independently of  each other.  The
correlation does not hold for short inner periods $P_S < 30$\,d.

Tidal interaction in multiple stars (KCTF) shortens the inner periods,
leading  to  the  excess  of  sub-systems with  periods  of  few  days
\citep{Fabrycky07}.   There is  a  strong  peak at  $\log  P_S =  0.25
... 0.75$ in  the distribution of inner periods  in the {\em complete}
sample of triple stars. However, it can be a selection effect produced
by  the eclipsing  systems  which are  discovered  at long  distances,
because this  peak almost disappears for nearby  triples within 50\,pc
(Fig.~\ref{fig:plpshist}).  The  distribution of the  inner periods in
the homogeneous radial-velocity survey of triple stars \citep{TS02} is
nearly flat out to $P_S \sim 60$\,d.  So, KCTF can produce only a mild
depletion at $P_S \sim 10$\,d but cannot explain the whole gap.

Note  that the period  distribution in  the volume-limited  samples of
dwarf binaries is a rising function in the $\log P$ interval from 0 to
3,  with  a   possible  dip  near  $\log  P   \sim  2$  \citep[Fig.~10
in][]{Halbwachs03}.  The  rise is  even steeper for  non-triple (pure)
binaries,  because  all SBs  with  $P<  3$\,d  are triple,  while  the
proportion of triples decreases  at longer periods \citep{Tok06}.  The
obvious  conclusion  is that  {\em  the  migration mechanism  produces
shorter orbital periods in triple stars than in pure binaries.}
It is then plausible that  the gap in the inner-period distribution in
triple  and quadruple  stars is  real and  is caused  by  migration to
shorter periods, rather than by the observational selection.

The distributions of the {\em outer} periods of quadruples and triples
are different.  All quadruples except  two have $\log P_L > 3.8$ ($P_L
> 17$\,y).   Moreover, there  is a  concentration of  points  near the
lower limit  $\log P_L \sim 4$,  not seen in the  triple systems where
the distribution  of $P_L$ is  smooth and extends to  shorter periods.
It is shown below that tight quadruples with $\log P_L < 4.5$ are also
distinguished by  larger outer mass  ratios $q_{L4}$ and  more similar
inner  periods, so the  reality of  this special  group of  objects is
based on  more than just outer  periods.  They are  further studied in
Appendix~\ref{sec:tight}.

The  quadruple system  with the  shortest  outer period  of 355\,d  is
VW~LMi, it combines a 0.5-day  eclipsing pair with a 8-day binary, all
composed  of dwarf  stars  \citep{VWLMi}.  The  triple  star with  the
shortest outer  period is $\lambda$~Tau:  $P_L= 33$\,d, $P_S  = 4$\,d,
$M_1 = 7.2 M_\odot$.

\subsection{Outer and inner mass ratios}
\label{sec:qq}

\begin{figure}
\includegraphics[width=8cm]{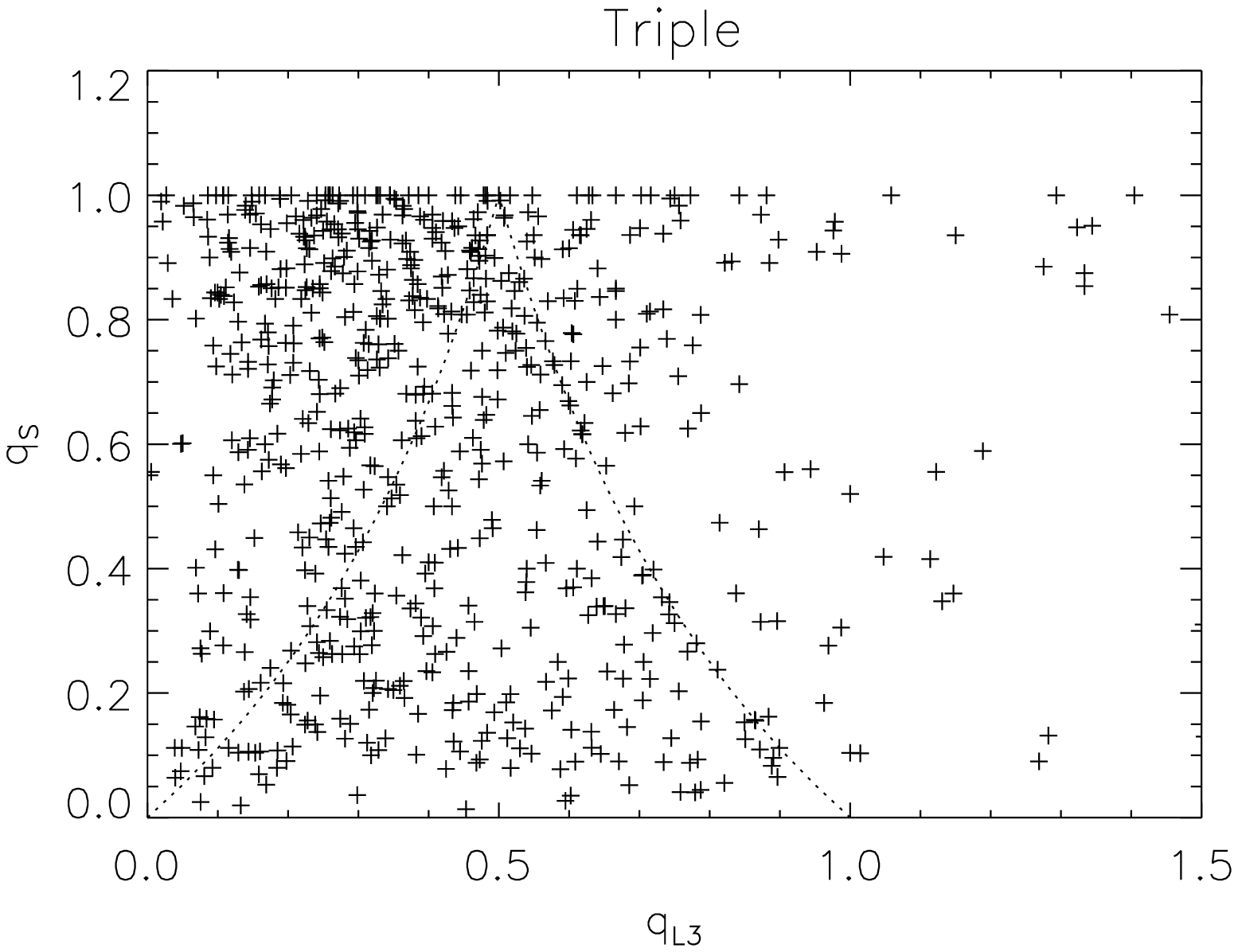} 
\includegraphics[width=8cm]{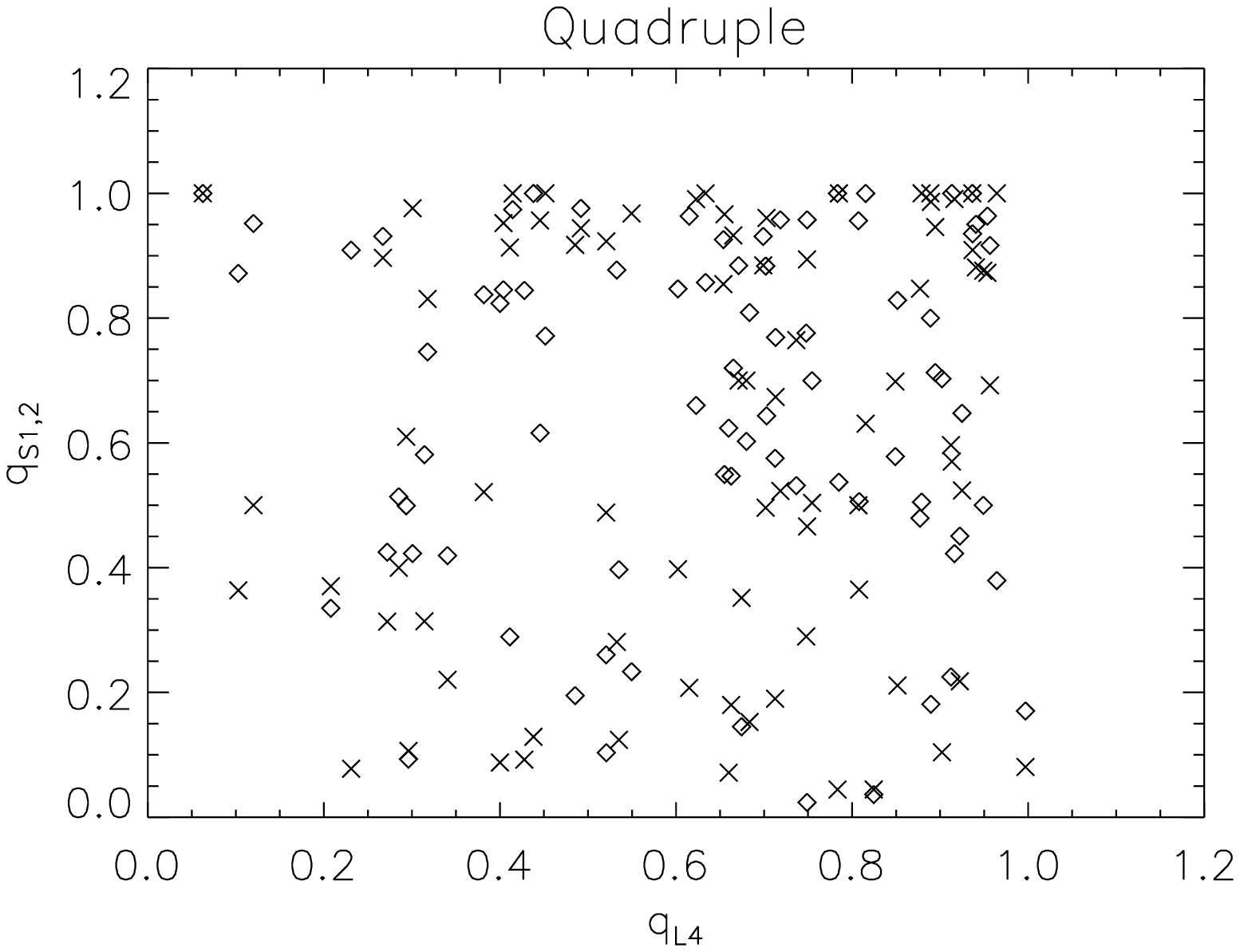} 
\caption{Comparison  between  inner  and  outer mass  ratios in
triples (top, the dotted  lines correspond to $M_3=M_2$ and $M_3=M_1$)
and quadruples (bottom, diamonds indicate $q_{S1}$, crosses $q_{S2}$).
\label{fig:q1q3}}
\end{figure}

It appears  that there is no  correlation between   inner and outer
mass    ratios    neither     in    triples    nor    in    quadruples
(Fig.~\ref{fig:q1q3}).   The  distributions of  $q_S$  in triples  and
$q_{S1},  q_{S2}$ in  quadruples look  quite similar.  Both  show some
excess  of  sub-systems  with  nearly identical  masses,  {\em  twins}
\citep{Halbwachs03,Lucy06,Sod07}.  

In triple  systems where  the mass of  the outer companion  equals the
mass of  an inner companion, $M_3  = M_1$ or  $M_3 = M_2$, there  is a
relation between inner  and outer mass ratios: $q_{L3}  = 1/(1 + q_S)$
and  $q_{L3}   =  q_S/(1  +  q_S)$,  respectively   (dotted  lines  in
Fig.~\ref{fig:q1q3}, top). There seems to be a concentration of points
along these lines, suggesting that  the phenomenon of twins may extend
to  the  outer  sub-systems  of  triple  stars  in  this  strange  and
un-explained way.

The  fraction of  triple systems  where  the outer  companion has  the
smallest mass, $M_3  < M_2$, is 332/724=46\% for  the whole sample and
71/184=39\% for triples within 50\,pc.  The median outer mass ratio is
$q_{L3} =  0.39$, 81\% of  triples have $q_{L3}>0.2$.   Thus, tertiary
companions tend to  have masses comparable with the  components of the
inner binary.  Triple systems resulting  from the N-body decay, on the
contrary, have only low $q_{L3}$ \citep[e.g. Fig. 4 in][]{DD03} and do
not  match real triples  in this  respect. Although  the observational
selection does favor multiples  with comparable masses, the difference
with the  simulations is  too strong to  be explained entirely  by the
selection.  The  survey of PMS stars by  \citet{Correia} confirms that
low-mass tertiaries are indeed rare, not just missed.

The distributions of the outer  mass ratios in quadruples and triples,
$q_{L4}$  and $q_{L3}$,  are  different. A  strong  tendency to  large
$q_{L4}$  is obvious,  97\%  of quadruples  having $q_{L4}>0.2$.   The
concentration   of    points   in   the   upper    right   corner   of
Fig.~\ref{fig:q1q3} corresponds to  the quadruples where all component
masses  are similar.   Simulations of  multiple stars  with components
selected independently are presented in Appendix~\ref{sec:imf} to show
how different they are from the real systems.

\subsection{Inner mass ratios vs. inner periods}
\label{sec:qsps}

\begin{figure}
\includegraphics[width=8cm]{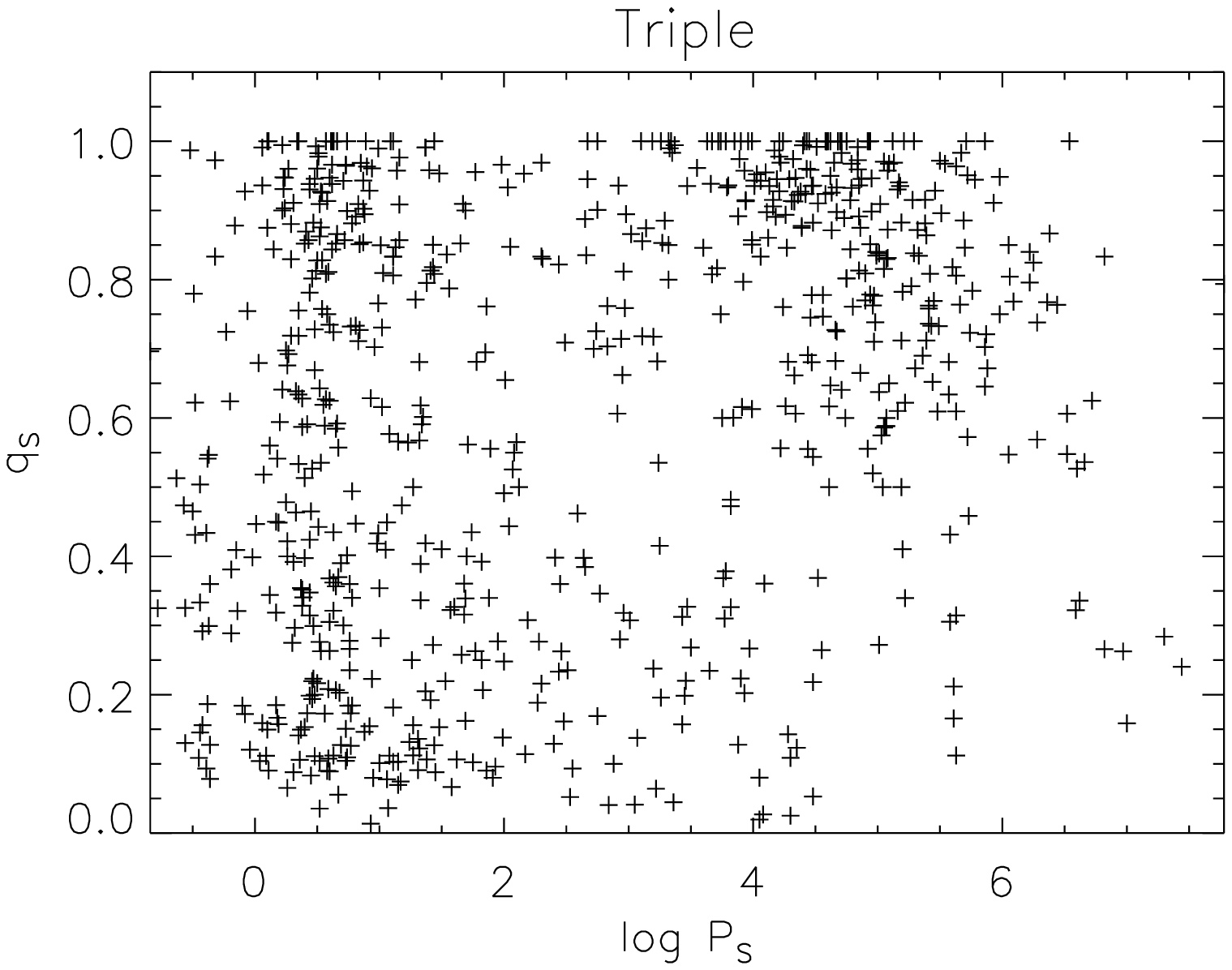} 
\includegraphics[width=8cm]{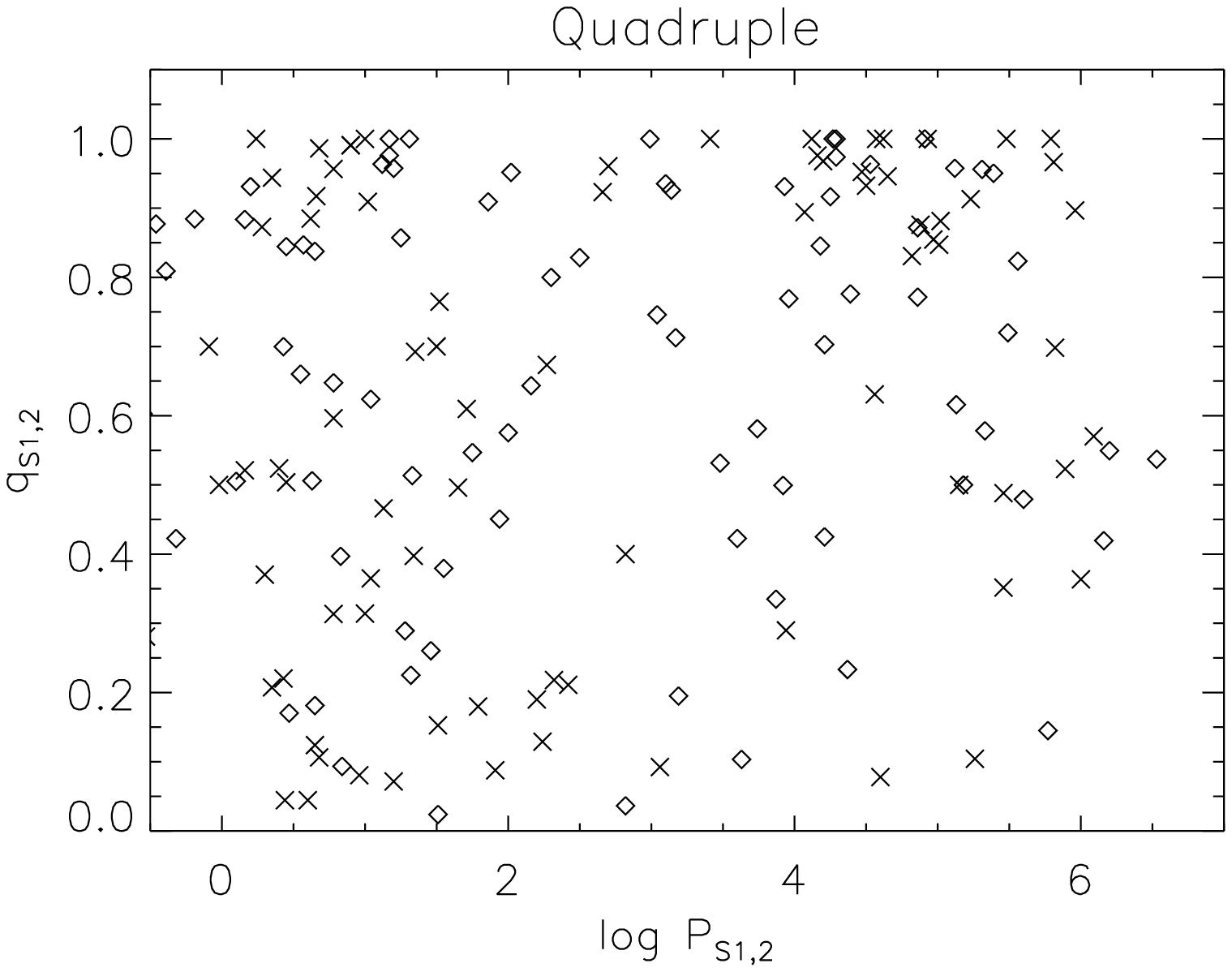} 
\caption{Mass  ratios vs. periods  $P_S$ in the  inner sub-systems  of
 triples (top) and
  quadruples (bottom, diamonds for the  primary pair and crosses for the
  secondary pair). 
\label{fig:qps}}
\end{figure}

Figure~\ref{fig:qps} presents a comparison of mass ratios with periods
for  the inner sub-systems.   The statistics  of triple  and quadruple
stars appear  to be  similar.  We note  the relative paucity  of inner
periods between 30 and $10^4$ days (the period gap, see above) and the
concentration of points  towards $q_S\sim 1$ (twins) on  both sides of
the gap. In  both samples the sub-systems with $\log P_S  > 4$ tend to
have more similar masses, $q_S>0.5$.   This trend may be caused by the
selection  (sub-systems with  low  $q$ are  not  discovered by  visual
techniques), as  illustrated by the GJ~225.2 system.   We checked that
the  distributions  of the  inner  and  outer  mass ratios  $q_S$  and
$q_{L3}'$ in nearby triple stars  with inner (resp.  outer) periods in
the  same range,  between $10^4$  and $10^6$  days,  are statistically
indistinguishable. On the other hand, sub-systems with short $P_S$ and
low $q_S$  could be  formed by an  alternative mechanism such  as disk
fragmentation.

\subsection{Outer mass ratios vs. outer periods}
\label{sec:qlpl}

\begin{figure}
\includegraphics[width=8cm]{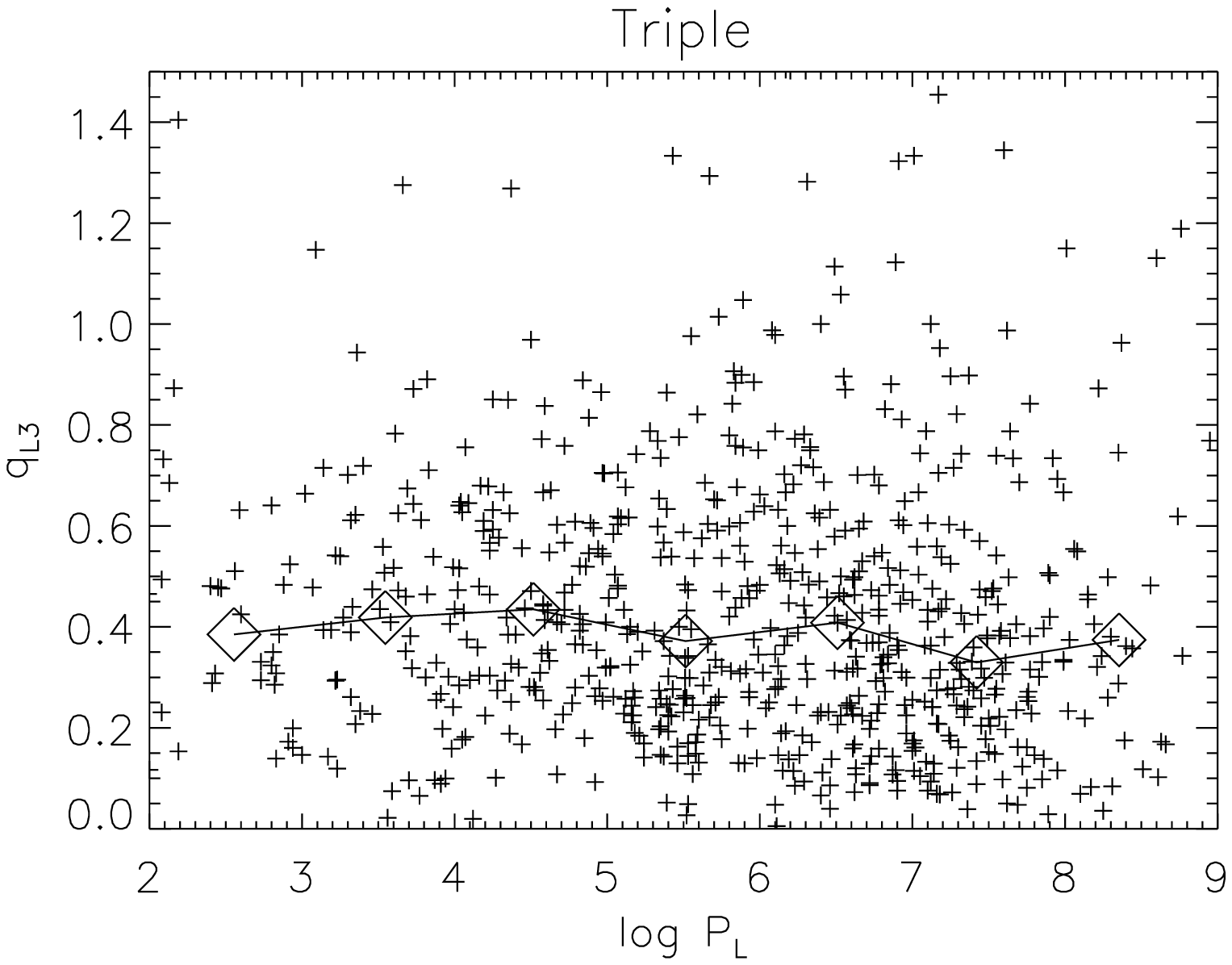} 
\includegraphics[width=8cm]{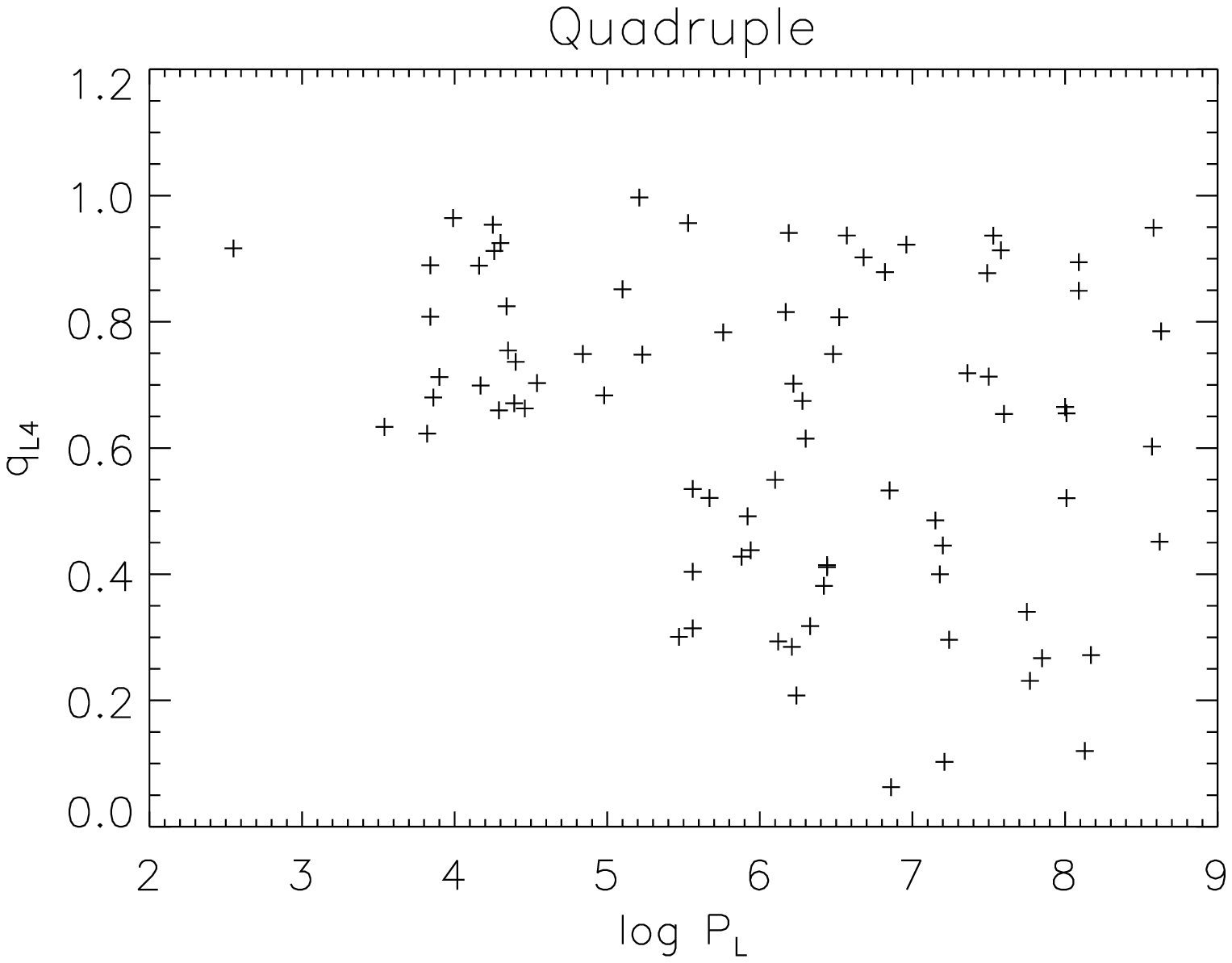} 
\caption{Mass ratios  vs. periods in  the outer sub-systems  of triple
 (top) and  quadruple (bottom) stars.   The diamonds show  median mass
 ratio for triples in each decade of the outer periods.
\label{fig:qpl}}
\end{figure}

Figure~\ref{fig:qpl} shows  the correlation of outer  mass ratios with
outer periods for triples  and quadruples.  These plots are strikingly
different.  In triple  stars, we see no dependence  of median $q_{L3}$
(diamonds) on  the outer period,  although the dispersion  of $q_{L3}$
seems  to increase  at  larger periods.   \citet{DD03} predicted  that
close inner  binaries with low-mass  tertiaries should be  common, but
there is  no correlation  between $q_{L3}$ and  $P_S$ to  support this
prediction. 


Note  the complete absence of  quadruples in the lower left corner with
$q_{L4}<0.6$ and $\log P_L<5.4$  and the relative abundance of triples
with  similar  parameters. This  difference  could  be  caused by  the
observational  selection.   Quadruples  with $q_{L4}<0.6$  can  escape
detection  because the  low-mass pair  is much  fainter than  the main
pair.  One such system with  estimated $q_{L4} = 0.63$ is $\chi$~Tau~B
where the light of the second binary is not detected at all, while its
binarity  is   inferred  from  the  total  mass   and  infrared  colour
\citep{Torres06}.   
%
%
Missed tight quadruples  with $q_{L4} < 0.6$ and  $P_L < 100$\,y would
be considered today as triples  with $q_{L3} < 0.6$.  We can establish
how  many  of such  low-$q_{L3}$  triples  are  actually quadruple  by
monitoring the  radial velocity of faint tertiary  components with the
detection technique of \citet{Dangelo06}.

\subsection{Orbital angular momenta}
\label{sec:mom}

\begin{figure}
\includegraphics[width=8cm]{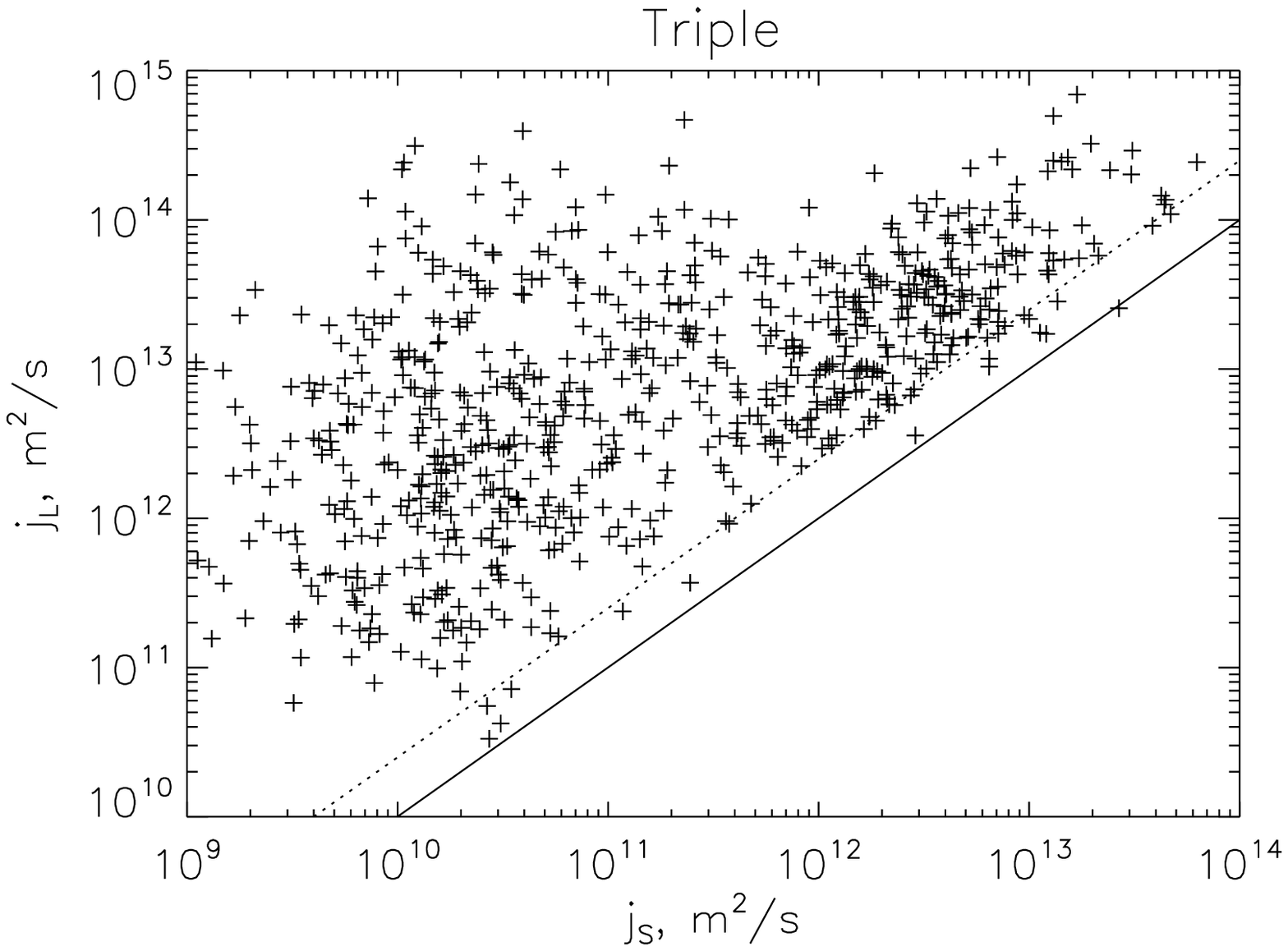} 
\includegraphics[width=8cm]{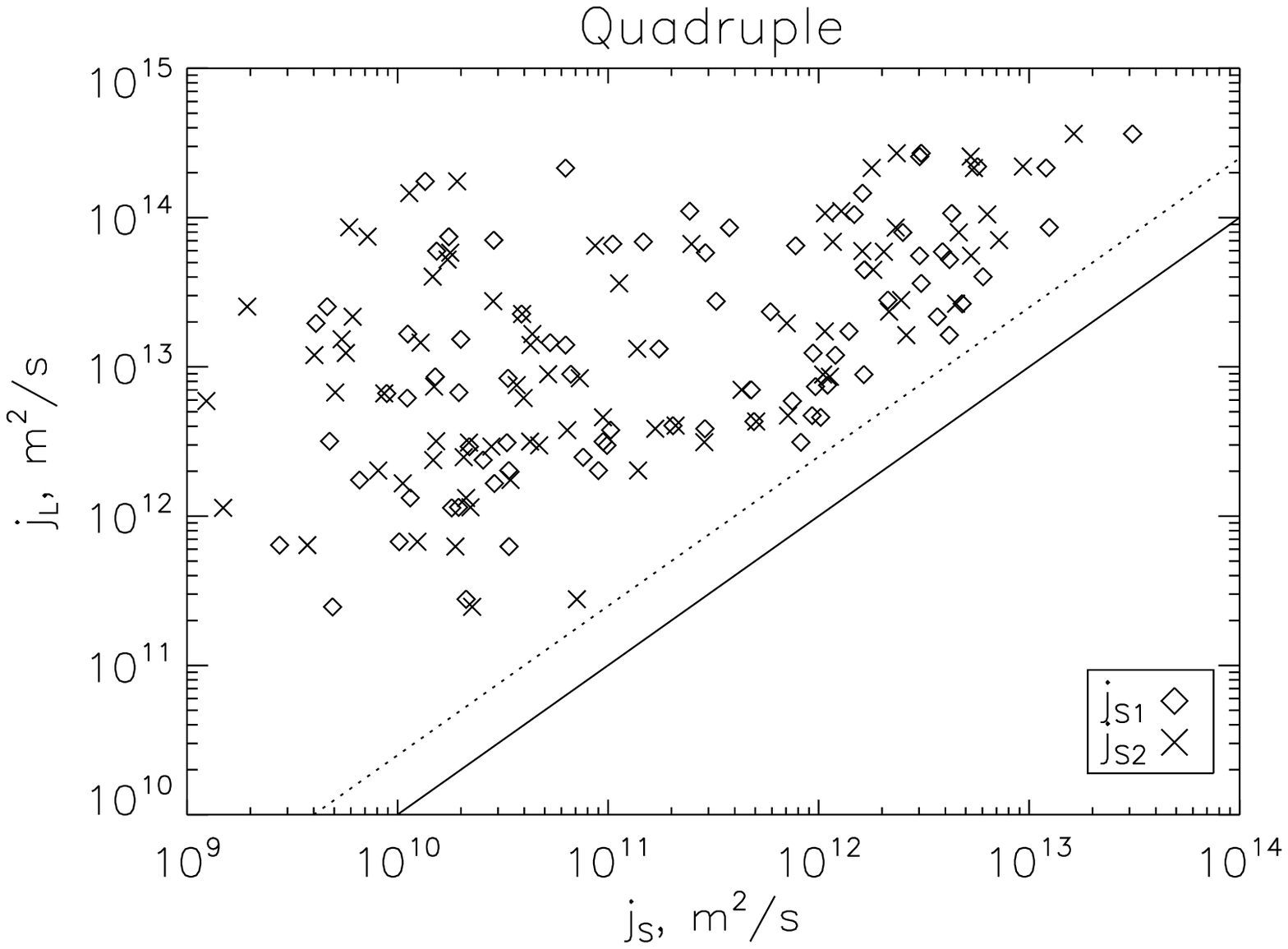} 
\caption{Specific orbital  angular  momentum  in  the  outer  sub-system  $j_L$
  vs. specific momentum  in the inner sub-system(s) $j_S$  in triple (top)
  and quadruple (bottom) stars. The  solid lines denote equality of the
  momenta, the dotted lines show $j_L/j_S = 2.5$.  
\label{fig:jljss}}
\end{figure}

Orbital angular momentum $J$ of a binary star equals 
\begin{equation}
J = \sqrt{ a (1 - e^2)}  M_1 M_2 \sqrt{G/M} ,
\label{eq:Jorb}
\end{equation}
where $M_1$ and  $M_2$ are the component's masses, $M =  M_1 + M_2$ is
the total mass,  $e$ is the eccentricity, $a$  is the semi-major axis,
and $G$ is the gravitational constant. In the following we neglect the
eccentricity because for a typical $e=0.5$ the factor $(1 - e^2)^{1/2}
= 0.87$ is  insignificant in comparison with the  large scatter of $J$
(there  are notable exceptions  like 41~Dra).   We calculate  $a$ from
periods  and  masses using  the  Kepler's  law.  The specific  angular
momentum $ j = J/M$ is  estimated for each sub-system.

Angular momentum  is strongly related  to orbital period,  so the
plots    in     Fig.~\ref{fig:jljss}    resemble    the     plots    in
Fig.~\ref{fig:plps}. The  correlation between inner  and outer periods
translates  into  the  correlation  between  the  momenta,  a  tighter
correlation for  wide multiples.  Considering that $J  \propto a^{1/2}
\propto P^{1/3}$,  the typical ratio  $J_L/J_S \sim 10$  translates to
$P_L/P_S \sim 10^3$.

Relative orientation  of the outer  and inner orbits  gives additional
clues to the  formation and dynamical evolution of  multiple stars. For
triples, it has been established  that the angular momentum vectors of
the inner and outer orbits do show some correlation, the average angle
between them, $\Phi$, being about $50^\circ$ instead of $90^\circ$ for
un-correlated spins \citep{ST02}. The  correlation of orbital spins is
stronger  at low $P_L/P_S$  ratios (close  to the  dynamical stability
limit) and disappears at large $P_L/P_S$.

There  are two  simple  quadruple stars,  88~Tau  and $\mu$~Ori,  with
complete elements  of both inner  and outer orbits known.   In 88~Tau,
the inner  orbits are not co-planar  to the outer orbit,  with $\Phi >
90^\circ$   (counter-rotation)    for   at   least    one   sub-system
\citep{Lane2007}.   Similarly,  the  Aab  sub-system in  $\mu$~Ori  is
counter-rotating, $\Phi  = 137^\circ \pm 8^\circ$, while  the orbit of
the  Bab  sub-system  is  nearly  perpendicular to  the  orbit  of  AB
\citep{MuOri}.  There  are other cases where the  apparent rotation of
the  resolved  inner and  outer  sub-systems  is opposite,  suggesting
non-aligned spins.  On the other  hand, some wide quadruples are close
to  alignment (e.g.   $\varepsilon$~Lyr, GG~Tau,  HD~98800).  Although
this evidence remains circumstantial, it hints on the same trend as in
the triples, i.e. the alignment of orbital spins for low $P_L/P_S$ and
mis-alignment  for large period  ratios.  A  dedicated interferometric
survey of inner  orbits is obviously needed to  increase the number of
multiple systems with known sense of relative rotation.

\subsection{Quadruple systems}
\label{sec:quad}

\begin{figure}
\includegraphics[width=8cm]{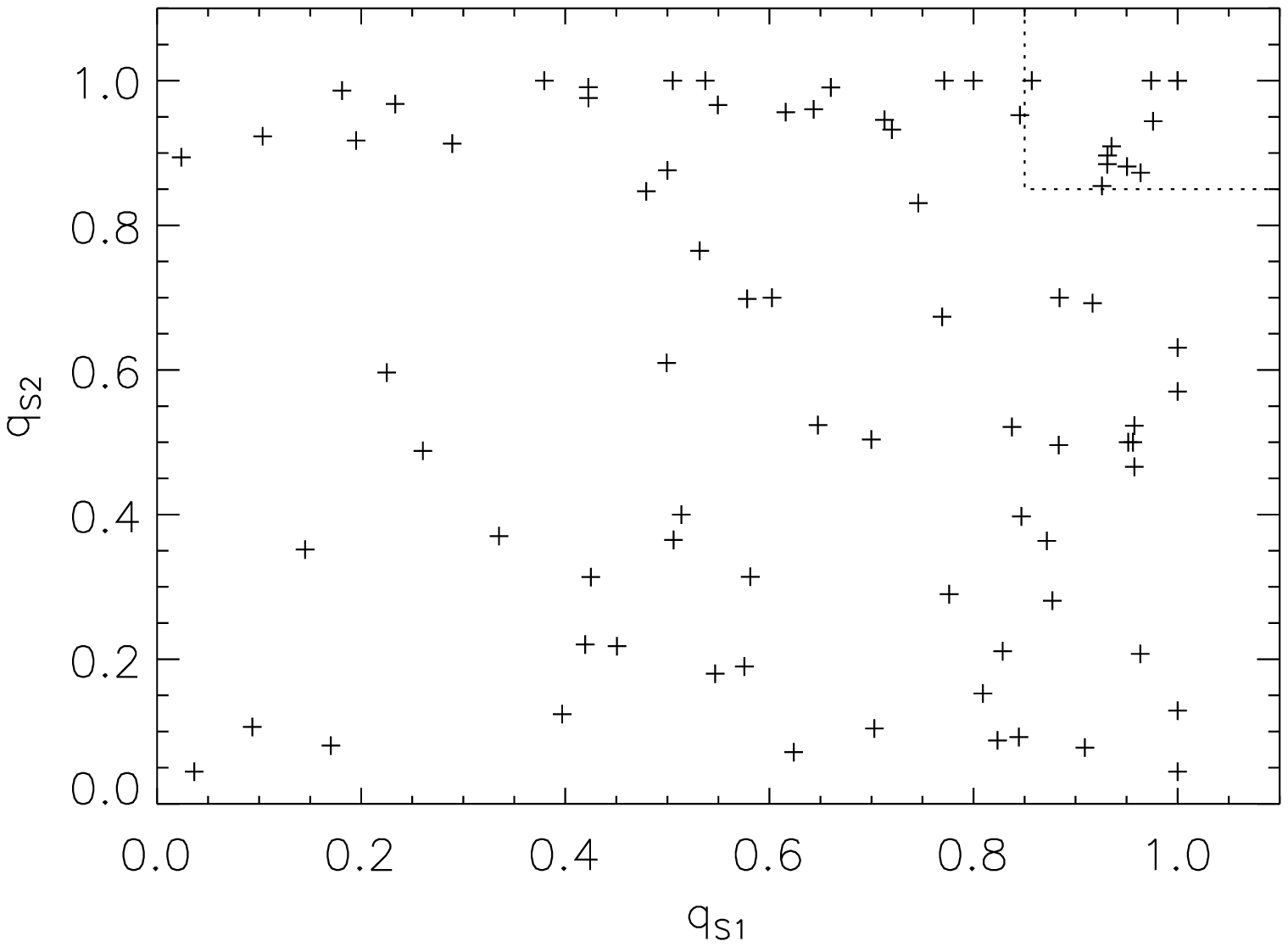}  
\includegraphics[width=8cm]{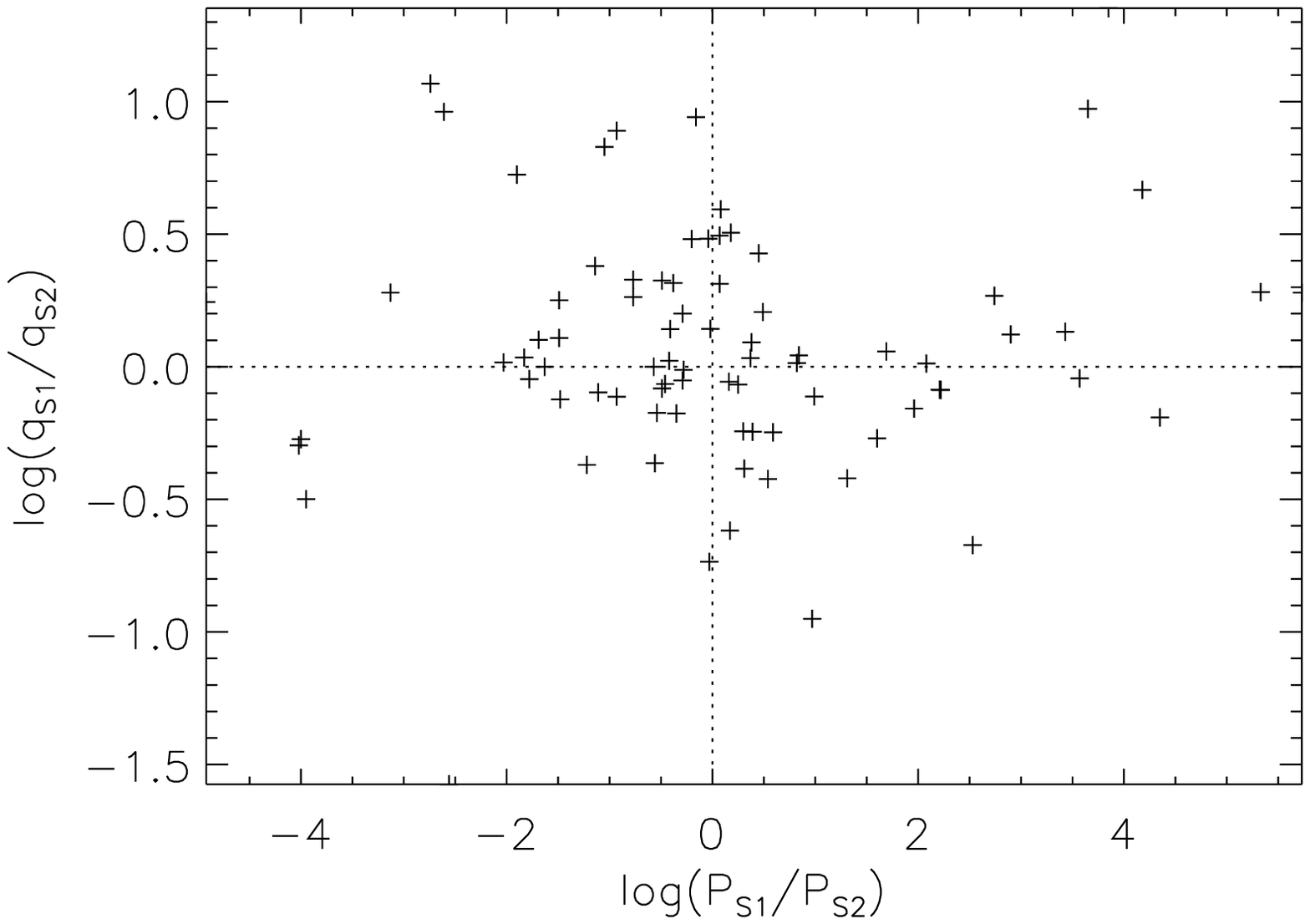} 
\caption{Top: Comparison  of the mass  ratios $q_{S1}$ and $q_{S2}$  in the
  inner  sub-systems of  quadruple stars.  The dotted  line  shows the
  region of double twins.
Bottom: Comparison between the  the inner periods and
the  inner mass ratios in quadruples. 
\label{fig:q1q2}}
\end{figure}

In this sub-section, we focus only on the quadruple stars and look for 
correlations between mass ratios and periods.
There is  no correlation  between the inner  mass ratios  $q_{S1}$ and
$q_{S2}$ (Fig.~\ref{fig:q1q2}).  However,  there is a concentration of
points towards  $q_{S2} \approx 1$,  and a less obvious  preference of
sub-systems with $q_{S1} \approx 1$.   There is a cluster of 11 points
in the upper right corner, $q_{S1}  > 0.85$, $q_{S2} > 0.85$ marked by
the  dotted   line  in  Fig.~\ref{fig:q1q2}.   If   the  points  were
distributed uniformly in the  $(q_{S1},q_{S2})$ plane between 0 and 1,
we would expect only 2.2\% in  this corner, while the actual number is
11/83=13\%. The  fraction of double  twins in the  simulated quadruple
systems with  independent component pairing is 0.25\%.   The family of
quadruples where  both sub-systems are twins ({\em  double twins} like
41+40~Dra)  thus appears  to be  distinct from  the majority  of other
quadruples.  BD$-22^\circ$~5866  is  yet  another  low-mass  quadruple
composed of  two twins \citep{Shkolnik}.  The  ``definition'' of twins
$q>0.85$ is adopted here arbitrarily, guided by Fig.~\ref{fig:q1q2}.

Sub-systems with $q_{S1}>0.85$ or $q_{S2}>0.85$ are found at all inner
periods   (cf.    Fig.~\ref{fig:qps},   bottom).   Interestingly,   the
proportion of twins in the sub-systems  of low mass seems to be higher
than in the high-mass ones.  Similarly, there are more twins among the
secondary   (less   massive)  sub-systems.    These   trends  need   a
confirmation with larger samples.  The  origin of twin binaries is not
yet understood. They could result from the preferential accretion onto
the  secondary that  drives $q$  towards one  and, at  the  same time,
shortens the orbital period  \citep{Bate2000}.  However, the inner mass
ratios do  not correlate with the  inner periods (Fig.~\ref{fig:qps}).
Figure~\ref{fig:q1q2} demonstrates the lack of correlation between the
periods and mass ratios in  the two inner sub-systems belonging to the
same quadruple.  Therefore, {\em the inner periods and mass ratios are
determined by different processes.}

\begin{figure}
\includegraphics[width=8cm]{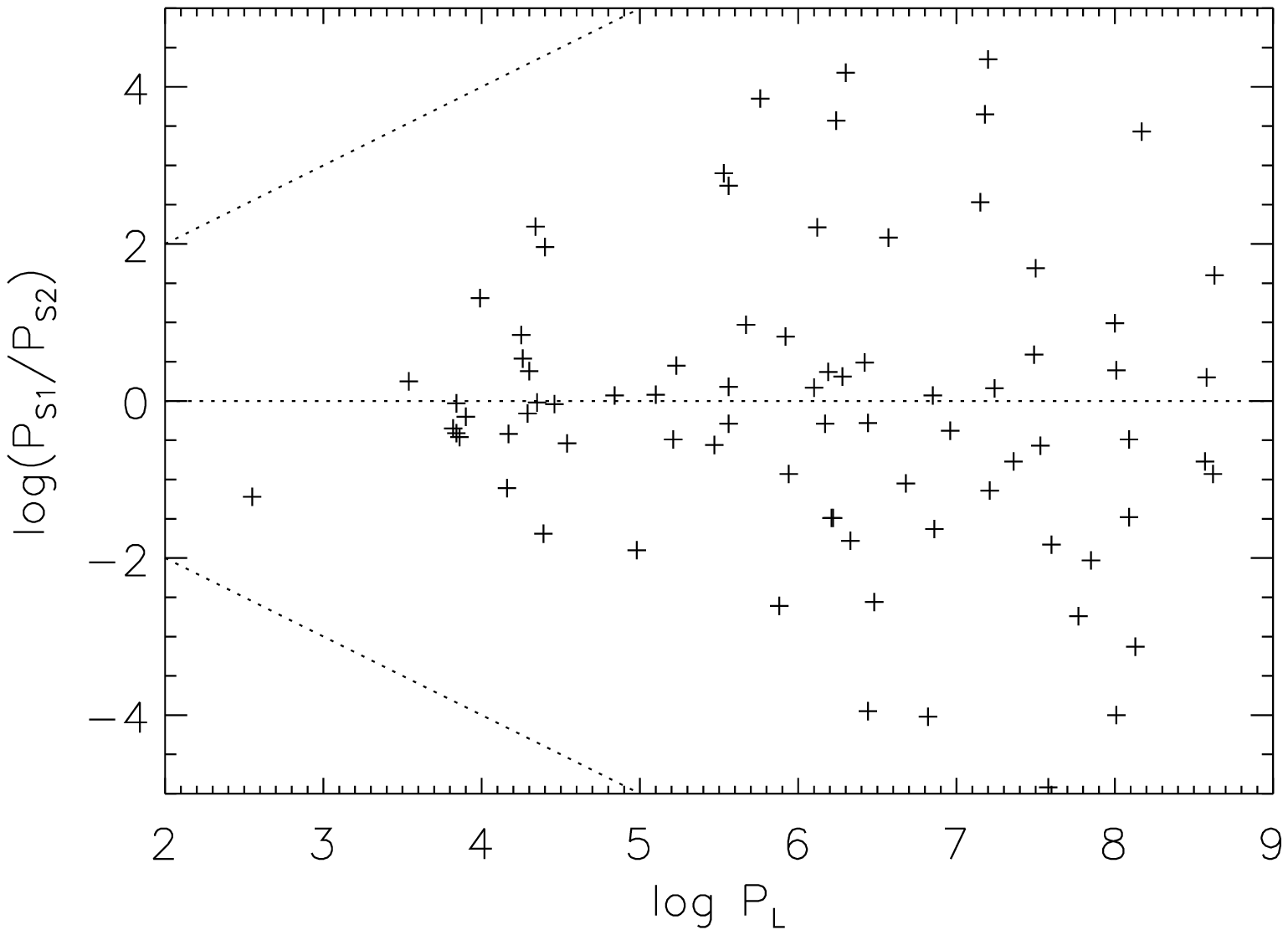}  
\includegraphics[width=8cm]{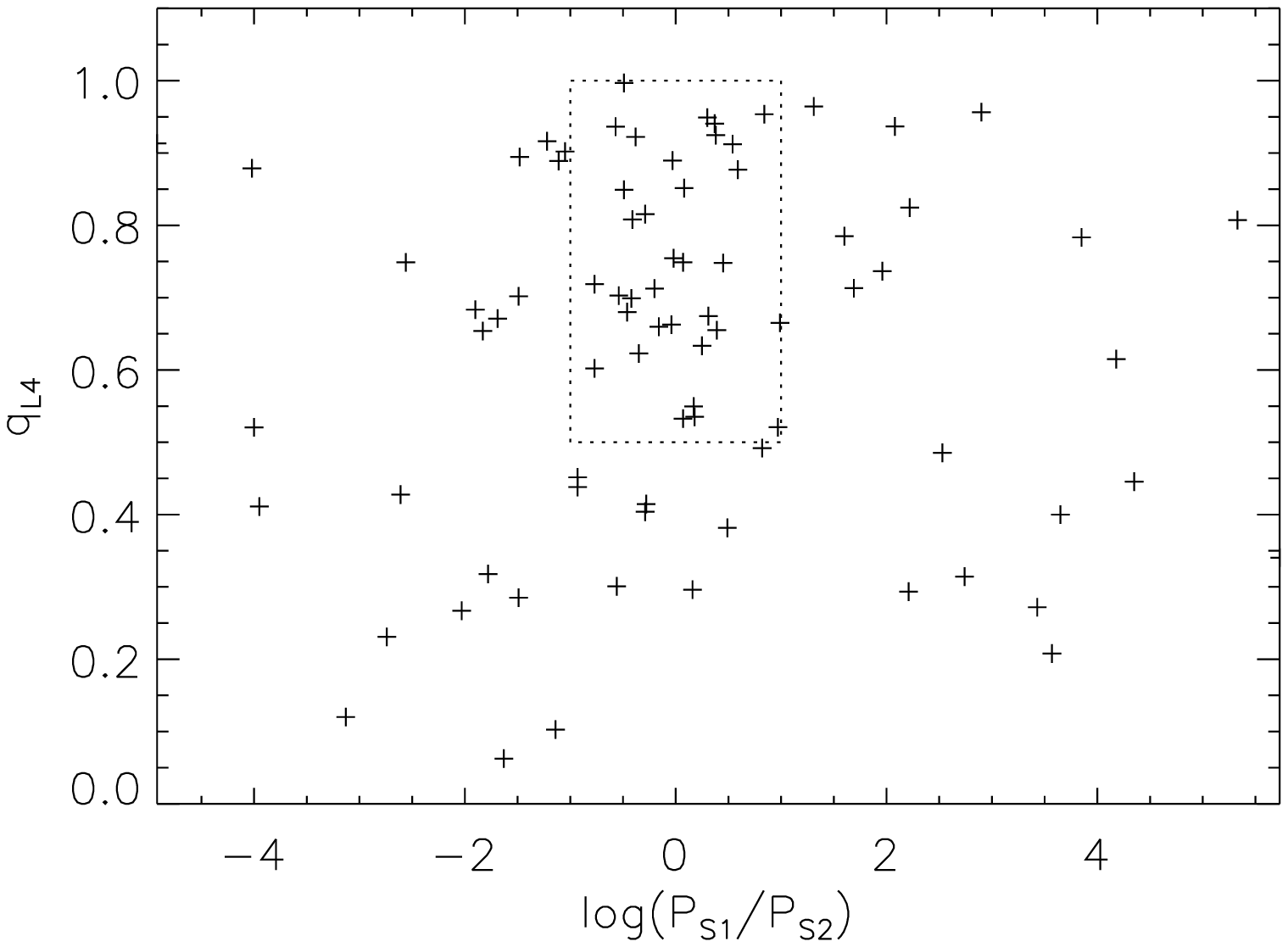}  
\caption{Top:  The ratio  of  the  inner periods  in
quadruple stars $  \log (P_{S,1}/P_{S,2})$ as a function  of the outer
period.
Bottom:  outer mass  ratio versus  inner  period ratio.  The region  of
$\varepsilon$~Lyr type quadruples with $| \log (P_{S1}/P_{S2} |<1$ and
$q_{L4}>0.5$ is delineated.  
\label{fig:dpspl}}
\end{figure}

The  points   in  Fig.~\ref{fig:q1q2}  (bottom)   concentrate  to  the
coordinate  origin, showing  that in  many quadruples  the  inner mass
ratios and inner periods are similar, as in the $\varepsilon$~Lyr.
The top  plot of Fig.~\ref{fig:dpspl} shows that  the period disparity
between the inner sub-systems in quadruple stars is less when $P_L$ is
short.  This  is expected because inner periods  are comprised between
few hours (contact binaries)  and the dynamical stability limit, $-0.7
< \log P_S < \log P_L  -0.7$.  At shorter $P_L$ this range is smaller,
hence  the  inner periods  are  forced to  be  more  similar, $|  \log
(P_{S,1}/P_{S,2})   |   <   \log   P_L$.    This   range,   shown   in
Fig.~\ref{fig:dpspl}  (top) by  the dotted  lines, is  wider  than the
distribution of the points by roughly 2\,dex. Therefore the similarity
of  the  inner  periods is  genuine  and  not  caused by  the  trivial
dynamical constraints.

The  numbers of quadruple  systems with  positive and  negative $|\log
(P_{S,1}/P_{S,2})|$ are statistically similar, 38 and 43 respectively,
so the periods  of the more and less massive  inner sub-systems do not
differ  systematically. The  migration process  that  determined these
periods should not depend on  the mass.  The inner periods are similar
to  within  1\,dex  in 42  systems,  i.e.   in  $52  \pm 8$\%  of  the
quadruples.  The relation  between the ratio of the  inner periods and
the  outer  mass ratio  is  further  explored in  Fig.~\ref{fig:dpspl}
(bottom).  As a rule, sub-systems with similar inner periods also have
similar total masses (larger $q_{L4}$).  There are 34 quadruples (42\%
of the sample) with $| \log (P_{S1}/P_{S2} |<1$ and $q_{L4}>0.5$, i.e.
resembling  $\varepsilon$~Lyr.    Conversely,  in  the   systems  with
dissimilar  masses (low $q_{L4}$),  the ratio  of inner  periods $\log
(P_{S,1}/P_{S,2})$  can take  large positive  or negative  values with
equal probability.

\section{Summary and discussion}
\label{sec:sum}

The properties of catalogued triple and quadruple systems are:
\begin{enumerate}
\item
  The distributions  of periods in the inner  sub-systems of quadruple
  and triple  stars are similar  and bimodal. They are  different from
  the period  distribution of pure binaries.  There  is no correlation
  between inner mass ratios and inner periods. 

\item
 The distributions of periods and mass ratios in the outer sub-systems
  of quadruple and triple stars  are different.  The median outer mass
  ratio in  triples is 0.39. It  does not depend on  the outer period,
  which has a smooth distribution.   In contrast, the outer periods of
  25\%  quadruples concentrate  in  the narrow  range  from 10\,yr  to
  100\,yr, all  these tight quadruples have $q_{L4}>  0.6$ (possibly a
  selection effect), while the periods of their two inner binaries are
  similar.

\item
 Quadruple  systems of  $\varepsilon$~Lyr type  with  comparable inner
 periods and component  masses are common.  In 42\%  of the quadruples
 the outer mass  ratios are above 0.5 and the  inner periods differ by
 less than 1\,dex.

\item
 The outer and inner mass ratios in triple and quadruple stars are
not mutually correlated. In 13\% of quadruples both inner mass ratios
are above 0.85 (double twins). 

\item
 The  values  of the  inner  and  outer  orbital angular  momenta  (or
periods) in  wide ($P_S  > 30$\,d) triple  and quadruple  systems show
some  correlation,  the  ratio  of outer-to-inner  periods  is  mostly
comprised  between 5  and $10^4$.  In  the systems  with small  period
ratios  (low  hierarchy)  the  directions  of the  orbital  spins  are
correlated, while in the systems with large ratios they are not.

\end{enumerate}


Statistical properties of multiple stars lead to several conclusions
regarding their formation mechanisms. 

{\bf 1.  N-body dynamics is  not the dominant process of multiple-star
formation.}  Triple  stars produced in  N-body decay (with  or without
accretion) have  distinctive properties: a  narrow period distribution
(width of  1--2\,dex), moderate period  ratios $P_L/P_S \sim  10$ (not
too far  from the dynamical  stability limit), high  eccentricities of
outer  orbits distributed  as  $f(e)=2e$ or  steeper  [cf.  Fig.~6  in
\citet{DD03} and  Fig.~6 in \citet{ST02}],  and low outer  mass ratios
$q_{L3}<0.2$.  Quadruple stars with 2+2 hierarchy are produced by this
process only  exceptionally. In contrast,  in real multiple  stars the
period  ratios can  be large,  the outer  eccentricities  are moderate
\citep[][see  also Table~\ref{tab:tight}]{Shatsky01,  Tok04},  and the
outer companions are often massive (81\% have $q_{L3}>0.2$). Dynamical
disruption of  small-N clusters leaves  too many single stars  and too
few binaries and multiples \citep{Goodwin05}.

The existence of  outer weakly bound components or  sub-systems, as in
the cases of $\alpha$~Gem and  88~Tau (at projected distances 1145 and
3500  AU, respectively),  speaks against  their dynamical  origin.  If
these  multiples were parts  of larger  clusters, the  stellar density
must be  low enough  for the survival  of the outer  components, hence
much too  low to  explain the dynamical  formation of the  close inner
sub-systems  where larger  binding energies  are  involved.

{\bf 2.   The rotationally-driven (cascade)  fragmentation scenario is
promising.} It  naturally produces  quadruple stars of  2+2 hierarchy.
As the motions in the pre-stellar cores are predominantly large-scale,
the net  angular momentum is sufficiently  high, preventing disruptive
N-body interactions between sub-systems.   The net angular momentum is
conserved  in  the  orbital  motion  of  the  outer  sub-system.   The
eccentricity of the  outer orbit in this scenario  should be moderate,
the system is hierarchical and dynamically stable.

Suppose that  two stellar pairs are  formed from the  fragments of one
core.  The initial separation between the components in each pair must
be of the  order of $10^2$ to $10^4$  AU.  The proto-binaries interact
with the surrounding gas, create spiral waves and transfer the orbital
angular  momentum  outwards, becoming  tighter  or  even merging  into
single stars  \citep[disk migration,][]{Krumholz07}.  The  merger must
not  happen  in the  case  of  quadruples,  however.  Each  sub-system
evolves independently of the  other system until its closest reservoir
of gas is exhausted.  As the directions of the net angular momentum in
each sub-condensation can differ, the orbital spins of the sub-systems
are not necessarily co-aligned with each other.

Accretion  of the  more distant  envelope  proceeds on  a longer  time
scale, when the  sub-systems already ``fall'' onto each  other and end
up forming a quadruple.  The matter from the outer envelope settles in
a  circum-quadruple  disk. The  result  resembles  a quadruple  system
GG~Tau, where the circum-quadruple disk slowly feeds the circum-binary
disk around the most massive  sub-system Aab. Some components may form
as fragments in accretion disks  around other, more massive stars. The
delayed formation of such components means that less mass is available
for  accretion,  possibly creating  binaries  with  low  $q$.  The  AB
quadruple in $\alpha$~Gem could be produced by such a process.

{\bf 3. Migration shortens both  inner and outer orbits.}\footnote{ 
As pointed out by the Referee, fragmentation at high densities may
form close binaries or multiples without migration, cf. \citet{Machida08}.}
It is likely
that there are several different types of migration.  Accretion onto a
binary and associated braking by the massive circum-binary disk is one
such  process confirmed  by hydrodynamical  simulations \citep{Bate02}.
Accretion usually increases the mass  ratio and can produce twins with
$q \sim 1$.   However, we find no correlation  between mass ratios and
periods  and infer  that  the dominant  migration mechanism  producing
close  binaries is  not accompanied  by the  accretion  of substantial
mass.  An interaction with circum-binary disk is a promising candidate
for  such  alternative  migration  mechanism.  Such  interaction  also
involves accretion, but possibly at a much reduced level not affecting
the  mass ratio.   It is  not clear  whether migration  can lead  to a
merger.  If  this is possible,  massive tertiary components  in triple
stars could be mergers in former 2+2 quadruples.

The gap  in the distribution of  the inner periods  can be tentatively
related   to   two   distinct   migration   mechanisms.    Presumably,
accretion-induced migration acts  at large separations, shortening the
inner (and sometimes outer)  periods to $\sim 30$\,y (separation $\sim
10$~AU) and  creating a  broad maximum in  the period  distribution at
$\log  P_S \sim  5$.  The  second migration  mechanism  shortens inner
periods even further (mostly below 30\,d) and produces the gap, but it
does  not affect the  mass ratio.  The second  migration must  be less
efficient in  pure binaries, so  their periods are longer  compared to
the   inner  sub-systems  of   multiple  stars   \citep{Tok06}.   This
explanation of the period gap is still highly hypothetical.

In the case of tight quadruples, a significant shrinking of both inner
and outer  orbits must  have happened.  Most  likely, the  outer orbit
shrinks  by accretion from  the   envelope, increasing  the outer
mass  ratio.  This  could  explain  high outer  mass  ratios of  tight
quadruples  (Fig.~\ref{fig:qpl})   and  matches  the   simulations  of
\citet{DD04}.   The gas accreted  by the  inner binaries  inherits its
angular momentum from the outer  orbit, so the similarity of the inner
periods in tight quadruples is  a natural consequence of such process.
We  also expect  the alignment  of  inner and  outer angular  momentum
vectors, which is  the case in some, but not  all triple and quadruple
stars.  Accretion of even a  small amount of gas can efficiently brake
the binary if  the angular momentum of the  gas is grossly mis-aligned
with respect  to the  binary, as must  happen in quadruple  and triple
stars  with  mis-aligned   (e.g.  counter-rotating)  inner  and  outer
systems.  A tight  quadruple with  mis-aligned inner  orbits,  such as
88~Tau, could  result from this  process.  It is noteworthy  that both
triples and quadruples with large  $P_L/P_S$ ratios show a tendency to
mis-aligned orbital spins.

To preserve  dynamical stability,  migration in the  inner sub-systems
must happen before or proceed faster than in the outer orbits.  In the
hydrodynamical  simulations of  \citet{Bate02}  some multiple  systems
become dynamically unstable because their outer orbits shrink too fast
due  to  accretion  and  disk  braking.  We  may  be  witnessing  such
situation in GG~Tau which interacts with its circum-quadruple disk and
is now  close to the  stability limit \citep{Beust06}.  A  fraction of
present-day triple stars may have  been produced by dynamical decay of
2+2 quadruples.  The triple star  $\lambda$~Tau with $P_L = 33$\,d and
$P_L/P_S  =  8$ could  result  from  the  outer-orbit migration  which
stopped just before the dynamical breakup.

{\bf 4.  Kozai  cycles with tidal friction (KCTF)  is not the dominant
process in producing close  sub-systems.} For long initial $P_S$, this
mechanism ``turns on'' in a very narrow range of relative inclinations
near $90^\circ$. Moreover, at large $P_L/P_S$ ratios as commonly found
in triple systems  the Kozai effect becomes too  weak and is perturbed
by other factors such as relativistic precession.  Hence, a population
of inner systems with sufficiently short $P_S$ must be created by some
other process before the KCTF  can further shorten some periods to few
days  \citep{Fabrycky07}. This  said, inner  binaries with  periods of
$\sim 1$\,d  must be produced  late, when the components  have already
settled on the MS, either by KCTF or by magnetic braking.

\begin{figure}
\includegraphics[width=8cm]{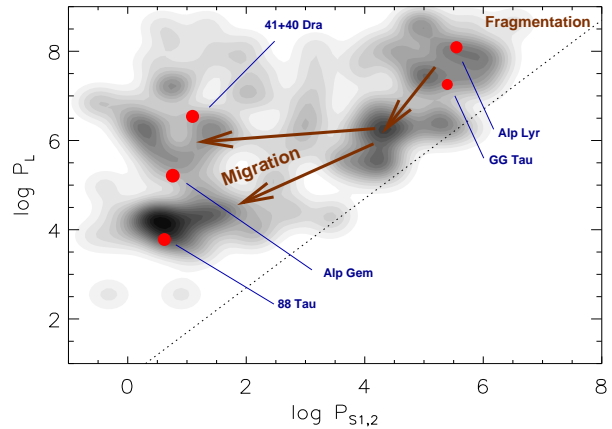} 
\caption{Schematic  evolution  of   quadruple  stars.  The  grey-scale
  background  shows the  smoothed  version  of the  $P_L  - P_S$  plot
  for quadruple systems (Fig.~\ref{fig:plps}).
\label{fig:evol}}
\end{figure}

Figure~\ref{fig:evol}  illustrates   the  scenario  of  quadruple-star
formation with a $P_L - P_S$ diagram where some individual systems are
located.  The area  of long inner and outer  periods can be identified
with   the   systems  that   did   not   evolve  significantly   after
fragmentation. Otherwise, the points migrate to the lower left in this
plot, faster in $P_S$ than in $P_L$ to stay always above the dynamical
stability limit.

Cascade fragmentation with subsequent orbit  evolution is a promising 
explanation for   the  origin  of  multiple   stars,  including  tight
quadruples.   This scenario  remains  very sketchy,  however, as  many
details are still not clear,  many alternatives remain to be explored.
From the  theoretical side, simulations  of accretion onto  a multiple
system   with  non-coplanar   orbits   are  needed.   Observationally,
establishing the unbiased statistics  of triple and quadruple stars in
well-defined samples, such as nearby dwarfs, will be a crucial step in
understanding the formation on these objects. We hope that the present
catalogue-based study will stimulate such projects.

\section*{Acknowledgements}

The author has discussed  the multiple-star statistics and the initial
draft  of this  paper with  M.~Sterzik, P.~Eggleton,  H.~Zinnecker and
other   coleagues.   Their   input   and  criticism   are   gratefully
acknowledged.

\appendix

\section{Tight quadruples}
\label{sec:tight}

A  group  of  {\em  tight   quadruples}  with  $\log  P_L  <  4.5$  is
distinguished in the  period-period plot (Fig.~\ref{fig:plps}). 
They also tend to have  comparable masses and periods of the inner
sub-systems. These properties are already obvious at $\log P_L < 5.4$,
so the  exact upper limit  $P_L$ to distinguish tight  quadruples from
the rest is not known.  We set this limit at $P_L = 100$\,y ($\log P_L
< 4.56$) and study the properties of such quadruples in more detail.

Table~\ref{tab:tight}  lists  all  20  known quadruples  with  $P_L  <
100$\,y:  the  WDS codes,  alternate  identifications, outer  periods,
apparent semi-major  axes (or separations) of the  outer orbits $a_L$,
their eccentricities $e_L$, and  primary masses $M_{1,1}$.  Only 3 out
of  20  quadruples  still  have  no  computed  outer  orbits  (missing
eccentricities in Table~\ref{tab:tight}).  In 7 systems there are 5 or
6 known components, like  in 88~Tau. This information is complementary
to the main Table~3.

The separations in  the outer orbits are sub-arcsecond  in 18 systems,
so the study  of the inner orbits did not  rely on spatially resolving
the outer pairs,  but rather on the spectroscopy  of components in the
combined  light.  Short  outer  periods can  be  easily discovered  by
spectroscopy, therefore the sharp drop  in the number of quadruples at
$\log P_L < 3.5$ must be genuine, not a selection effect.  T.~Pribulla
(private  communication) confirms  that in  a large  sample  of triple
stars discovered  spectroscopically the outer periods  are mostly long
and the tertiary companions are  mostly single, so the quadruple
VW~LMi with $P_L =355$\,d stands out as a unique system.

\begin{table}
\center
\caption{Tight quadruples}
\label{tab:tight}
\begin{tabular}{ll c c c c }
\hline
WDS(2000) & Name &  $P_L$     & $a_L$,  & $e_L$ & $M_{1,1}$  \\ 
          &      &  yr        &  $''$   &       & $M_\odot$   \\
\hline
00247-2709 &GJ 2005    & 40   &    1.07   & --   & 0.10 \\  
04078+6220 &SZ Cam     & 61   &    0.061  & 0.83 & 15.4 \\    
04226+2538 &$\chi$ Tau B& 9.5  &    0.084  & 0.35 & 1.19 \\    
04356+1010 &88 Tau     & 18.0 &    0.241  & 0.08 & 2.06 \\    
04518+1339 &VB 124     & 95   &    0.73   & 0.60 & 1.29 \\    
06024+0939 &$\mu$ Ori  & 18.8 &    0.276  & 0.76 & 2.76 \\    
08119-4844 &AO Vel     & 41   &    0.038  & 0.29 & 3.63 \\    
08285-0231 &HR 3337    & 55   &    0.283  & 0.43 & 2.10 \\    
10017+1725 &XY Leo     & 19.6 &    0.148  & 0.11 & 0.78 \\    
10444-6000 &QZ Car     & 50   &    0.025  & --   & 40.0 \\    
11029+3025 &VW LMi     & 0.97 &    0.013  & 0.14 & 1.68 \\    
11182+3131 &$\zeta$ UMa& 60   &    2.54   & 0.41 & 1.10 \\    
11395-6524 &GT Mus     & 80   &    0.217  & --   & 2.78 \\    
14593+4649 &ET Boo     & 67   &    0.196  & 0.55 & 1.21 \\    
15420+0027 &HIP 76892  & 54   &    0.158  & 0.72 & 2.10 \\    
16058+1041 &NQ Ser     & 20   &    0.230  & --   & 0.83 \\    
20432+5707 &HR 7490    & 21   &    0.047  & 0.36 & 10.6 \\    
21424+4105 &77 Cyg     & 27   &    0.153  & 0.34 & 2.98 \\    
23019+4220 &o And      & 68   &    0.277  & 0.48 & 5.85 \\    
23304+3050 &HIP 116017 & 48   &    0.189  & 0.43 & 1.10 \\    
\hline
\end{tabular}
\end{table}

\section{Mass ratios with independent pairing}
\label{sec:imf}

\begin{figure}
\includegraphics[width=8cm]{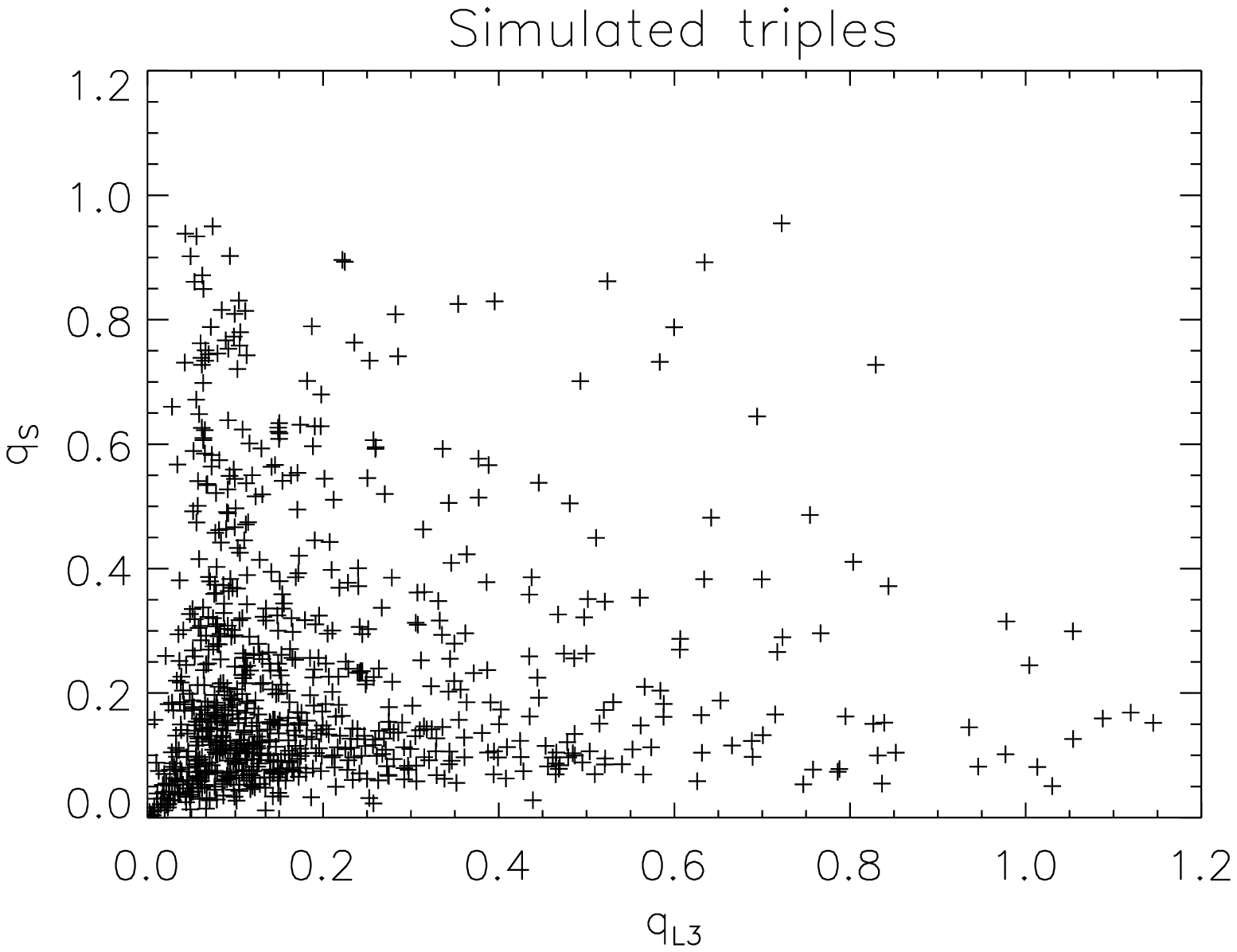} 
\includegraphics[width=8cm]{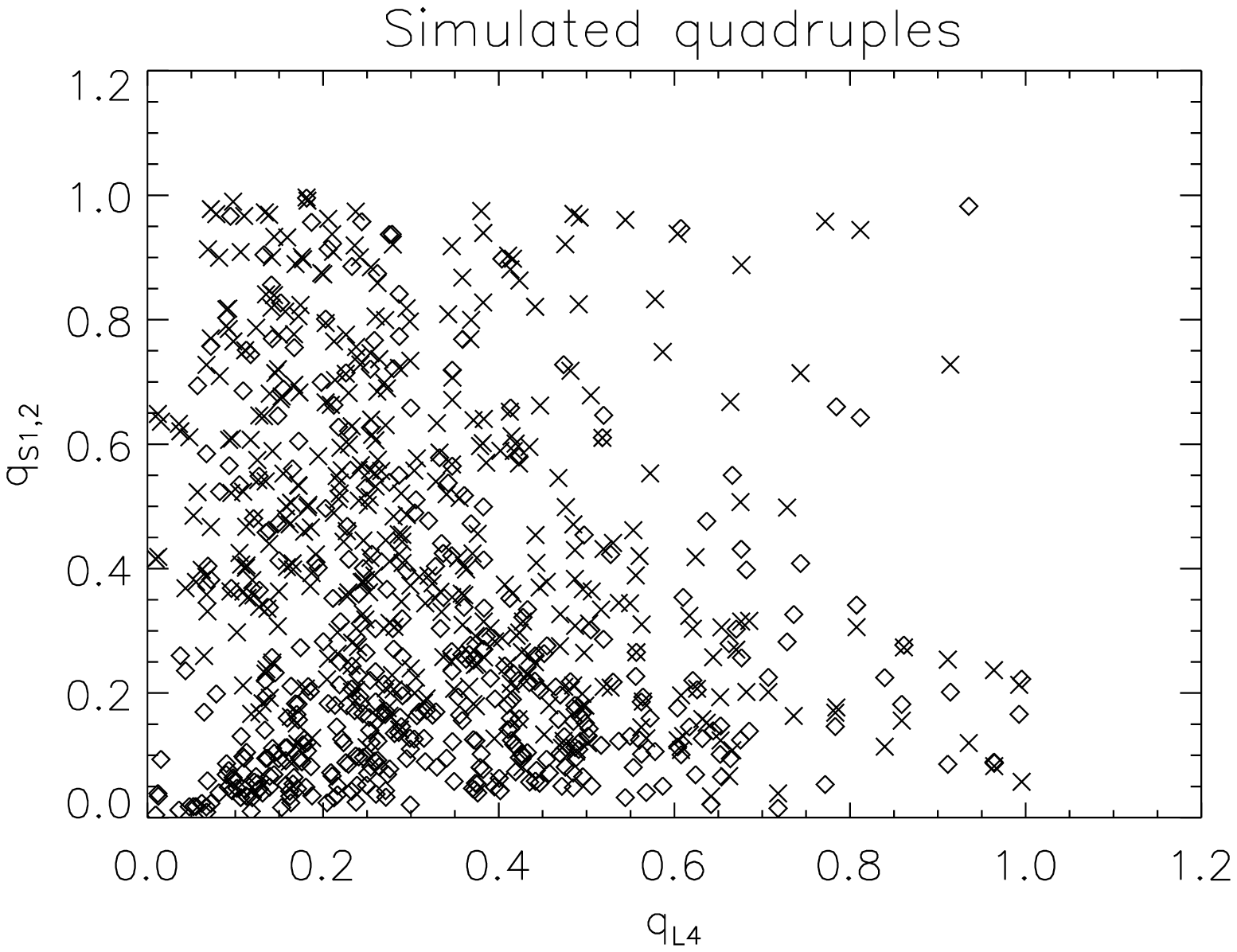} 
\caption{Mass ratios in 865  simulated triples (top) and 463 simulated
quadruples  (bottom) with independent  component masses  selected from
the IMF.  Compare with Fig.~\ref{fig:q1q3}.
\label{fig:q1q3sim}}
\end{figure}

Is  is  established  that the  mass  ratios  in  binary stars  do  not
correspond  to  the random  selection  of  component  masses from  the
initial  mass  function  (IMF) \citep[e.g.][]{Halbwachs03,Kou07},  the
component masses rather  tend to be comparable.  The  same tendency is
expected in the multiple  systems.  Nevertheless, it is instructive to
simulate  mass  ratio  distributions  resulting from  the  independent
pairing  and to compare  them with  the data.   We used  the numerical
recipe of \citet{Fisher04} to  generate the masses.  Only triples with
$M_1>M_{\rm min}$ and quadruples with  $M_{1,1} > M_{\rm min}$, with $
M_{\rm  min} =  0.8  M_\odot$, were  retained  from a  larger pool  of
simulated systems dominated by small masses. This is done to mimic our
samples,  although the  distribution  of simulated  primary masses  is
more strongly concentrated  towards $ M_{\rm  min}$ than the  actual masses
(Fig.~\ref{fig:m1hist}).  The  plots of the simulated  mass ratios are
shown  in  Fig.~\ref{fig:q1q3sim}.   The  difference with  the  actual
distributions is too strong to  be attributed solely to selection.  If
the masses in  the real multiple stars were  indeed independent and we
were missing a huge number of multiples with low mass ratios, it would
mean  that the  actual frequency  of  multiple systems  is many  times
higher than observed now.  Systems with similar components are rare in
the simulated multiples.  The median simulated mass ratios are $q_{L3}
=  0.13$, $q_{L4}  = 0.29$.   A  loose correlation  between the  small
values of  $q_S$ and $q_{L3}$ seen  in the Figure is  explained by the
threshold imposed on the primary mass, while the other components most
frequently have similarly low masses near the IMF peak.


\bsp

\label{lastpage}


\begin{thebibliography}{99}


\bibitem[\protect\citeauthoryear{Bate}{2000}]{Bate2000}
Bate M., 2000, MNRAS, 314, 33 

\bibitem[\protect\citeauthoryear{Bate, Bonnell, \& Bromm}{2002}]{Bate02}
Bate M.R., Bonnell I.A., \& Bromm V., 2002, MNRAS, 336, 705


\bibitem[\protect\citeauthoryear{Beust \& Dutrey}{2006}]{Beust06}
Beust H. \& Dutrey A., 2006, A\&A, 446, 137

\bibitem[\protect\citeauthoryear{Boden et al.}{2005}]{Boden05}
Boden A.F., Sargent A.I., Akeson R.L., Carpenter J.M., 
Torres G., Latham D.W., Soderblom D.R., Nelan E., Franz O., \&
Wasserman L.H., 
2005, ApJ, 635, 442


\bibitem[\protect\citeauthoryear{Bodenheimer}{1978}]{Bodenheimer78}
Bodenheimer P., 1978, ApJ, 224,  488


\bibitem[\protect\citeauthoryear{Bonnell}{2001}]{Bonnell2000}
Bonnell I., 2001, in Zinnecker H. \& Mathieu R.D., eds, The formation of binary
stars. Proc. IAU Symp. 200. ASP Conf. Ser.

\bibitem[\protect\citeauthoryear{Correia et al.}{2006}]{Correia}
Correia S., Zinnecker H., Ratzka Th., \& Sterzik M.F., 
2006, A\&A, 459, 909

\bibitem[\protect\citeauthoryear{D'Angelo, van Kerkwijk, \& Rucinski}{2006}]{Dangelo06}
D'Angelo C., van Kerkwijk M.H., \& Rucinski S.M.,
 2006, AJ, 132, 650

\bibitem[\protect\citeauthoryear{Delgado-Donate,  Clarke,  \& Bate}{2003}]{DD03}
Delgado-Donate E.J., Clarke C.J., \& Bate M.,
2003, MNRAS, 342,  926

\bibitem[\protect\citeauthoryear{Delgado-Donate et al.}{2004}]{DD04}  
Delgado-Donate  E.J.,  Clarke  C.J., Bate  M.R., \&  Hodgkin
S.T., 2004, MNRAS, 351, 617

\bibitem[\protect\citeauthoryear{Eggenberger et al.}{2007}]{Egg07}
Eggenberger A., Udry S., Chauvin G.,
Beuzit J.-L., Lagrange A.-M., S\'egransan D., \& Mayor M.,
2007, A\&A, 474, 273

\bibitem[\protect\citeauthoryear{Eggleton}{2006}]{Eggleton}
Eggleton P., 2006, 
Evolutionary Processes in Binary and Multiple Stars.
Cambridge Univ. Press, Cambridge, UK


\bibitem[\protect\citeauthoryear{Fabrycky \& Tremaine}{2007}]{Fabrycky07}
Fabrycky D. \& Tremaine S., 2007, ApJ, 669, 1298

\bibitem[\protect\citeauthoryear{Fekel}{1981}]{Fekel81}
Fekel F.C., 1981, ApJ, 246, 879

\bibitem[\protect\citeauthoryear{Fisher}{2004}]{Fisher04}
Fisher R.T., 2004, ApJ, 600, 769

\bibitem[\protect\citeauthoryear{Goodwin,  Whitworth,  \& Ward-Thompson}{2004}]{Goodwin04}
Goodwin S.P., Whitworth A.P., \& Ward-Thompson D., 2004, A\&A, 414, 633

\bibitem[\protect\citeauthoryear{Goodwin \& Kroupa}{2005}]{Goodwin05}
Goodwin S.P. \& Kroupa P.,
2005, A\&A, 439, 565

\bibitem[\protect\citeauthoryear{Halbwachs et al.}{2003}]{Halbwachs03}
Halbwachs J.L., Mayor M., Udry S., \& Arenou F., 2003, A\&A, 397, 159

\bibitem[\protect\citeauthoryear{Kouwenhoven et al.}{2007}]{Kou07}
Kouwenhoven M.B.M.,  
Brown A.G.A., Portegies Zwart S.F., \& Kaper L.,  2007, A\&A, 474, 77

\bibitem[\protect\citeauthoryear{Krumholz, Klein,  \& McKee}{2007}]{Krumholz07}
Krumholz M.R., Klein R.I., \& McKee C.F., 2007, ApJ, 656, 959

\bibitem[\protect\citeauthoryear{Lane et al.}{2007}]{Lane2007}
Lane B.F., Muterspaugh M.W., Fekel F.C., Williamson M., Konacki M.,
Burke B.F., Colacita M.M., Kulkarni S., \& Shao M., 
2007, ApJ, 669, 1209


\bibitem[\protect\citeauthoryear{Lucy}{2006}]{Lucy06}
Lucy L.B., 2006, A\&A, 457, 629

\bibitem[\protect\citeauthoryear{Machida et al.}{2008}]{Machida08}
Machida M.N., Tomisaka K., Matsimoto T., \& Inutsuka S., 2008,
ApJ, 677, 327

\bibitem[\protect\citeauthoryear{Mason et al.}{2001}]{WDS}
Mason B.D., Wycoff G.L., Hartkopf W.I., Douglass G.G., \& Worley
C.E.,  2001, AJ, 122, 3466


\bibitem[\protect\citeauthoryear{Muterspaugh et al.}{2008}]{MuOri}
Muterspaugh M.W., Lane B.F., Fekel F.C., Konacki M., 
Burke B.F., Kulkarni S.R., Colavita M.M.,  Shao M., \&
Wiktorowicz S.J. 2008, AJ, 135, 766

\bibitem[\protect\citeauthoryear{Nidever et al.}{2002}]{Nidever02}
Nidever D.L., Marcy G.W.,  Butler R.P.,  Fischer D.A., \& Vogt
S.S., 2002,  AJ, 141, 503

\bibitem[\protect\citeauthoryear{Pepli\'nski, Artymowicz, \& Mellema}{2008}]{Mig3}
Pepli\'nski A., Artymowicz P., \& Mellema G., 2008, MNRAS, 386, 179

\bibitem[\protect\citeauthoryear{Pribulla et al.}{2006}]{VWLMi} 
Pribulla T., 
Rucinski S.M., Lu W., Mochnacki S.W., 
 Conidis G., Blake R.M., DeBond H., Thompson J.R., Pych W., Ogloza W.,
 \& Siwak M., 2006, AJ, 132, 769
et al.

\bibitem[\protect\citeauthoryear{Reipurth \& Aspin}{2004}]{Reipurth04}
Reipurth Bo \& Aspin C., 2004, ApJ, 608, L65

\bibitem[\protect\citeauthoryear{Ribas \& Miralda-Escud\'e}{2007}]{Ribas07}
Ribas I. \& Miralda-Escud\'e J., 2007, A\&A, 779



\bibitem[\protect\citeauthoryear{Shatsky}{2001}]{Shatsky01}
Shatsky N., 2001, A\&A, 380, 238


\bibitem[\protect\citeauthoryear{Shkolnik et al.}{2008}]{Shkolnik}
Shkolnik E., Liu M., Reid I.N., Hebb L., Vameron A.C., Torres C., \& 
Wilson D.M., 2008, preprint (arXiv:0805.0312)

\bibitem[\protect\citeauthoryear{S\"oderhjelm}{2007}]{Sod07}
S\"oderhjelm S., 2007, A\&A, 463, 683

\bibitem[\protect\citeauthoryear{Stamatellos, Hubber, \& Whitworth}{2007}]{Stamatellos07}
Stamatellos D., Hubber D.A., \& Whitworth A.P., 2007, 
MNRAS, 382, L30

\bibitem[\protect\citeauthoryear{Sterzik \& Durisen}{1998}]{SD98}
Sterzik M.F., \& Durisen R.H., 1998, A\&A, 339, 95

\bibitem[\protect\citeauthoryear{Sterzik, Durisen, \& Zinnecker}{2003}]{Sterzik03}
Sterzik M., Durisen R.H., \& Zinnecker H.
2003, A\&A, 411, 91 

\bibitem[\protect\citeauthoryear{Sterzik \&  Tokovinin}{2002}]{ST02}
Sterzik M. \&  Tokovinin A., 2002, A\&A,  384, 1030


\bibitem[\protect\citeauthoryear{Tokovinin}{1997}]{MSC}
Tokovinin A.A., 1997, A\&AS,  124,  75
See also http://www.ctio.noao.edu/\~{}atokovin/stars/index.php 

\bibitem[\protect\citeauthoryear{Tokovinin}{2004}]{Tok04}
Tokovinin A., 2004,  in: Proc. IAU Coll. 191, ed. C. Scarfe, C. Allen. 
Rev. Mex. Astron. Astrof., Conf. Ser., 21,  7

\bibitem[\protect\citeauthoryear{Tokovinin}{2008}]{Tok08}
Tokovinin A., 2008, in: Multiple stars across the H-R
diagram. Springer, p. 37

\bibitem[\protect\citeauthoryear{Tokovinin et al.}{2003}]{41Dra}
Tokovinin A., Balega Y.Y., Pluzhnik E.A., Shatsky N.I., Gorynya N.A.,
Weigelt G., 2003, A\&A,  409,  245

\bibitem[\protect\citeauthoryear{Tokovinin et al.}{2005}]{GJ2252}
Tokovinin A., Kiyaeva O., Sterzik M., Orlov V., Rubinov A., \&
Zhuchkov R., 2005, A\&A, 441, 695

\bibitem[\protect\citeauthoryear{Tokovinin \& Smekhov}{2002}]{TS02}
Tokovinin A.A. \& Smekhov M.G., 2002, A\&A,  382,  118

\bibitem[\protect\citeauthoryear{Tokovinin et al.}{2006}]{Tok06}
Tokovinin A., Thomas S., Sterzik M., \& Udry S., 2006, A\&A,  450,  68

\bibitem[\protect\citeauthoryear{Torres}{2006}]{Torres06}
Torres G., 2006, AJ,  131, 1702

\bibitem[\protect\citeauthoryear{Zinnecker \& Mathieu}{2001}]{IAU200}
Zinnecker H. \& Mathieu R.D., eds, The formation of binary
stars. Proc. IAU Symp. 200. ASP Conf. Ser., 2001.


\end{thebibliography}
\end{document}